\documentclass
[aps,prl,reprint,superscriptaddress,amsmath,amssymb,showpacs]{revtex4-1}
\usepackage{amsmath,amssymb,mathrsfs}
\usepackage[dvips]{graphicx}
\usepackage{psfrag,latexsym}
\usepackage{colordvi}

\date{\today}                  

\begin{document}

\title{Nonequilibrium Kondo model: Crossover
from weak to strong coupling}

\author{Mikhail Pletyukhov}
\email[Email: ]{pletmikh@physik.rwth-aachen.de}
\affiliation{Institut f\"ur Theorie der Statistischen Physik, RWTH Aachen, 
52056 Aachen, Germany and JARA - Fundamentals of Future Information Technology}
\author{Herbert Schoeller}
\affiliation{Institut f\"ur Theorie der Statistischen Physik, RWTH Aachen, 
52056 Aachen, Germany and JARA - Fundamentals of Future Information Technology}

\begin{abstract}
We analyze the nonequilibrium Kondo model at finite voltage and
temperature by using a new formulation of the real-time renormalization group 
method with the
Laplace variable as the flow parameter. We evaluate the energy-dependent
spin relaxation rate and nonlinear conductance, and derive an
approximate form for the universal line shape for the latter 
in the whole crossover regime from weak to strong coupling. The results are shown to
agree well with exact methods in equilibrium, Fermi-liquid 
theory, weak-coupling expansions, and recent experiments. 
For the transient spin dynamics we find a 
universal exponential decay in the long-time limit along with 
a truncation-dependent pre-exponential power law. For multichannel models
a pure power-law decay typical for non-Fermi-liquid behaviour is predicted.

\end{abstract}


\maketitle
The solution of the nonequilibrium Kondo model is a basic unsolved problem
in condensed matter physics. In its elementary
version, the model consists of a spin ${1\over 2}$ coupled 
antiferromagnetically to two fermionic metallic reservoirs. It reveals
the Kondo effect of complete spin screening at low energies and plays an important role in bulk systems 
with magnetic impurities \cite{hewson}. In the Coulomb blockade regime of single-level 
quantum dots, the Kondo model can be realized via its relation to the symmetric 
Anderson impurity model \onlinecite{kondo_theo}. In this case the effective exchange coupling
is given by $J\sim{2\Delta\over\pi U}$, where $\Delta$ is the level broadening and
$U\sim D$ is the charging energy, defining the band width $2D$. 
This allows for a controlled experimental realization of the
Kondo model in nonequilibrium, where the Kondo effect has been observed via
a zero-bias anomaly \cite{kondo_exp_general}. In equilibrium, the Kondo model has
been solved by a Bethe ansatz \onlinecite{bethe_ansatz} and conformal field 
theory (CFT) \cite{cft}. Using the numerical renormalization group (NRG) method
\cite{costi_etal_94,weichsel,wilson_75} and a Bethe ansatz \cite{konik_etal_01_02}, the temperature dependent linear conductance $G(T)$ has been shown to be a universal function
of $T/T_K$, where $T_K\sim \,D e^{-1/(2J)}$ is the
Kondo temperature. Analytical results are available for low ($T\ll T_K$) 
and high ($T\gg T_K$) temperatures from Fermi-liquid theory and poor man 
scaling methods \cite{nozieres_74,oguri_JPSJ05,poor_man_scaling}. 
A challenging issue is the nonequilibrium properties at $T=0$, such as the
determination of the nonlinear conductance $G(V)$ as a function of bias 
voltage $V$
and the transient spin dynamics $\langle\underline{S}\rangle(t)$.
The weak-coupling regime $\text{max}\{V,1/t\}\gg T_K$ 
has been solved by improved poor man scaling methods using phenomenological decay rates 
\cite{rosch_etal_general,glazman_pustilnik_05,doyon_andrei_PRB06}, flow equations \cite{kehrein_general,hackl_etal_09},
and the real-time renormalization group (RTRG) method \cite{hs_review_EPJ09,hs_reininghaus_PRB09,pletyukhov_etal_PRL10}.
However, the analysis of universal properties in the strong coupling regime 
$V,1/t\lesssim T_K$ remains an open problem. 

In this Letter we propose an approximate solution of this problem by using a new formulation
of the RTRG method where the Laplace variable $E$ is used as the flow parameter (called \lq\lq $E$-flow''
in the following). Within this scheme the relaxation rates occurring in each renormalization group (RG) step are the 
full physical ones, and  their energy dependence appears to be crucial for the description of 
the crossover from weak to strong coupling. Applying the $E$-flow to the nonequilibrium 
Kondo model, we calculate the nonlinear conductance 
and observe that both the $T$ and $V$ dependence agree well with recent experiments 
\cite{kretinin_etal_11}. We show analytically that in
the scaling limit ($D\rightarrow\infty$, $J\rightarrow 0$, $T_K=\text{const}$) 
{\it both} the low and high energy properties are consistent with Fermi-liquid theory 
\cite{oguri_JPSJ05} and weak-coupling expansions, respectively. This distinguishes our 
approach from other studies of the nonequilibrium Anderson impurity model (see, {\it e.g.}, 
Refs.~\onlinecite{eckel_etal_NJP10} and \onlinecite{andergassen_etal_review10} for reviews), 
where the effective 
spin exchange coupling $J$ cannot be chosen arbitrarily small to achieve universality in the scaling 
limit. Also approximate calculations of $G(V)$ via the Bethe ansatz eigenstates \cite{konik_etal_01_02} cannot cover the crossover to the weak coupling regime.
We find that our $G(T)$ curve agrees within a few percent with the NRG one and that 
the nonlinear conductance $G(V)$ is almost independent of the truncation order. 

Finally we address the transient spin dynamics after an initial coupling of the spin to the bath.
In Ref.~\cite{pletyukhov_etal_PRL10} it has been predicted in the weak coupling regime that the 
long-time dynamics is always exponential, accompanied by pre-exponential power 
laws in higher orders. 
Scaling behavior has been observed within time-dependent NRG calculations \cite{anders_schiller_05_06} and exponential 
behavior was predicted at the Thoulouse point \cite{lobaskin_kehrein_PRB05} and in the
framework of integrable quantum field theories \cite{lesage_saleur_PRL98}. However, due to the
presence of branch-cut contributions the transient spin dynamics is still under debate (see
Ref.~\cite{spin_boson} for a discussion within the spin boson model). 
At $T=V=0$ we find a universal exponential decay 
in the long-time limit $t\gg 1/T_K$ with a rate $\Gamma^*\sim T_K$, and, in addition, a 
pre-exponential power-law behavior with a truncation-dependent exponent. In contrast, for 
multichannel Kondo models, we predict a pure power-law decay in the long time limit as a 
result of non-Fermi-liquid behaviour at low energies \cite{cft,multi_channel}.

\emph{Model.}---We consider a single spin $\frac12$ with spin $\underline{S}$, which is
coupled by an isotropic exchange $H_{ex}=J_0\underline{S}\cdot\underline{s}$ to the spins
$\underline{s}=\frac12\sum_{\alpha\alpha'\sigma \sigma' kk'}a^\dagger_{\alpha\sigma k}
\,\underline{\sigma}_{\sigma\sigma'}\,a_{\alpha'\sigma' k'}$
of two reservoirs labeled by $\alpha=L,R$. $\sigma=\downarrow,\uparrow$ denotes the
spin, $k$ is the state index and $\underline{\sigma}$ are the Pauli matrices.
The reservoirs are described by $H_{res}=\sum_{\alpha\sigma k}\epsilon_{\alpha k}a^\dagger_{\alpha\sigma k}a_{\alpha\sigma k}$ with
finite band width and are kept on chemical potentials $\mu_{L/R}=\pm V/2$ (we use units $e=\hbar=k_B=1$).

\emph{Liouvillian approach.}---Following Ref.~\onlinecite{hs_review_EPJ09}, we  describe the 
dynamics of the reduced density 
matrix $\rho(t)$ of the local spin by the effective von Neumann
equation $\dot{\rho}(t)=-i\int_{0}^t dt'L(t-t')\rho(t')$, which reads in 
Laplace space $\rho(E)=i/(E-L(E))\rho_0$ \cite{endnote_1}. At the initial time $t=0$ 
the system is assumed to factorize into an arbitrary local part $\rho_0$ and 
a grand canonical equilibrium part $\rho_L^{\text{eq}}\rho_R^{\text{eq}}$ for the reservoirs.
$L(E)$ is an effective Liouvillian, which results from integrating out the reservoir degrees
of freedom. Using spin conservation and rotational invariance for the isotropic Kondo model
we find that the Liouvillian can be parameterized by the spin relaxation rate $\Gamma(E)$ via the
nonvanishing matrix elements $L(E)_{ss,s's'}=-{i\over 2}ss'\Gamma(E)$ and $L(E)_{s,-s,s,-s}=-i\Gamma(E)$,
where $s=\uparrow,\downarrow\equiv \pm$. The Liouvillian has a zero eigenvalue corresponding to
the stationary state and three degenerate eigenvalues $-i\Gamma(E)$ describing the spin dynamics via
\begin{equation}
\label{eq:S(E)}
\langle \underline{S}\rangle(E)\,=\,{i\over E+i\Gamma(E)}\,\langle \underline{S}\rangle(t=0)\,,
\end{equation}
from which the real-time dynamics can be obtained via the inverse Laplace transform.
Analogously, one can also derive an equation for the current 
$I_\alpha(E)=-i\text{Tr}\Sigma_\alpha(E)\rho(E)$ flowing 
into the reservoir $\alpha$, where $\Sigma_\alpha(E)_{ss,s's'}=i\Gamma_\alpha(E)$ are the
nonvanishing matrix elements of the current kernel. Defining 
$G(E)=\pi{\partial\over\partial V}\Gamma_L(E)$, the stationary conductance in units 
of $G_0=2e^2/h$ is given by $G=G(0)$. 

\emph{RG formalism.}---To calculate $\Gamma(E)$ and $G(E)$ we use a different formulation
of the RTRG method than the one proposed in Ref.~\onlinecite{hs_review_EPJ09}. Instead of introducing an 
artificial cutoff $\Lambda$ on the imaginary frequency axis, we use the physical Laplace variable 
$E$ itself as the flow parameter. This means that we differentiate the full diagrammatic series for the
effective kernels and vertices w.r.t. $E$ and obtain a self-consistent set of equations in
terms of effective quantities by an appropriate resummation of all diagrams (see details in the 
Supplementary Material \cite{SM}). It turns out that the scaling limit
is well-defined if two $E$-derivatives are taken for the kernels, whereas a single 
derivative is sufficient for the vertices. The exact RG equations are finally truncated systematically 
by considering all diagrams up to a certain order in the effective vertices but keeping the full
propagator between the vertices. The reliability of our approach in the strong coupling regime is tested
by comparing the results in second and third order truncation. These truncation schemes are essentially 
different from conventional $1$- and $2$-loop treatments, since we consider the full propagator
between the vertices and since we use the same truncation scheme for the rate $\Gamma(E)$ including
its full energy dependence. We illustrate this by showing the RG equations for $T=V=0$ 
(the full equations at finite $T,V$ can be found in the Supplementary Material \cite{SM}). Neglecting all terms of $O(J^4)$, we 
obtain with $E=i\Lambda$
\begin{equation}
\label{eq:rg}
d^2_\Lambda \Gamma = -{4J^2\over \Lambda+\Gamma}\quad,\quad
d_\Lambda J = -{2J^2(1+ZJ)\over \Lambda+\Gamma}\quad,
\end{equation}
where $Z={1\over 1+{d\Gamma\over d\Lambda}}$ denotes the $Z$-factor. With $\lambda=\Lambda+\Gamma$
the RG equation for the conductance reads $d_\Lambda G=-{3\pi^2\over 2\lambda}J_I K$, 
which is coupled to two other vertices with the RG equations $d_\Lambda J_I= -2J_I J/\lambda$ and 
$d_\Lambda K=-4KJ(1-ZJ)/\lambda$. With $\tilde{J}=ZJ$ we can write Eq.~(\ref{eq:rg}) in the form
$d_\lambda \tilde{J}=-2\tilde{J}(1-\tilde{J})/\lambda$ which gives the invariant
$T_K=(\Lambda+\Gamma)\sqrt{\tilde{J}\over 1-\tilde{J}}e^{-{1\over 2\tilde{J}}}$,
defining the Kondo temperature. In contrast, the standard $2$-loop Kondo temperature is
obtained by neglecting the relaxation rate $\Gamma$ leading to the poor man scaling solution 
$J_p(\Lambda)$, defined by $T_K=\Lambda\sqrt{J_p\over 1-J_p}e^{-{1\over 2J_p}}$.
Thus our approach provides a microscopic foundation for the physically intuitive picture 
of the relaxation rate cutting off the RG flow, including its full energy dependence. 
As an important consequence we find in our approach that the $2$-loop 
fixed point at $\Lambda=0$ is shifted to the negative value $\Lambda=-\Gamma^*<0$. 
As shown below this is fundamentally related to the exponential relaxation of the spin 
dynamics. Moreover, a technical advantage is that the point $E=i\Lambda=0$,
defining the stationary value of all physical quantities, is well separated from the fixed point. 
This allows for a well-defined expansion around this point leading to the correct Fermi-liquid relations (see below). 

The initial conditions at high energies $\Lambda_0\gg T_K$ are given by 
$J=J_0$, $Z=1$, $K=2J_0^2$, $J_I=J_0$ and $G={3\pi^2\over 4}J_0^2$ \cite{hs_review_EPJ09}.
In the scaling limit, this leads to the solution $Z=(1-\tilde{J})^2$ and $G={3\pi^2\over 4}\tilde{J}^2$.
For $\Gamma$ a special procedure is needed since it is proportional to $\Lambda$ in the weak coupling 
regime $\Gamma=2\Lambda (J_p + O(J_p^2))$. This leads to a truncation order independent instability for 
its value in the regime $\Lambda\ll T_K$. We solve this
problem by complementing our approach with the exactly known result of unitary conductance 
$G(E=0)={3\pi^2\over 4}\tilde{J}^2(E=0)=1$, which is a consequence of the 
Friedel sum rule \cite{hewson}. This fixes the values 
$\bar{J}\equiv\tilde{J}(0)={2\over \pi\sqrt{3}}\approx 0.37$, $\bar{\Gamma}\equiv\Gamma(0)=
\sqrt{1-\bar{J}\over\bar{J}}e^{{1\over 2\bar{J}}}\,T_K\approx 5.11 \,T_K$, and
$\bar{Z}=(1-\bar{J})^2\approx 0.40$, and the RG equations (\ref{eq:rg}) can be solved numerically starting
from $\Lambda=0$. The $T=V=0$ results for all quantities at large $\Lambda_0\gg T_K$ are then used as 
initial condition to solve the RG equations at finite $T,V$ using an initial value
$\Lambda_0\gg T,V,T_K$.
\begin{figure}[t]
  \centering
  \includegraphics[width=65mm,clip=true]{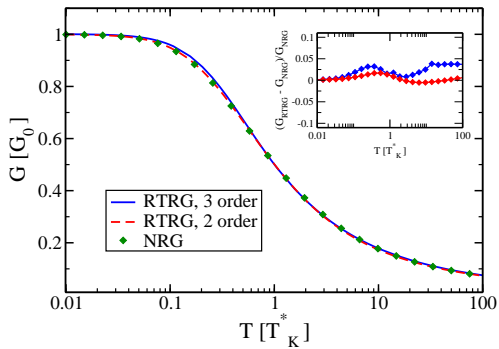}
  \caption{(Color online)
    Main panel: $G(T)$ in second (red dashed line) and third (blue full line) order 
    truncation as function of $T/T_K^*$, defined by $G(T=T_K^*)=1/2$, compared to the 
    NRG results (green diamonds).
    Inset: $(G_{\text{RTRG}}-G_{\text{NRG}})/G_{\text{NRG}}$ as function of $T/T_K^*$
    in second (lower line) and third (upper line) order truncation.
  }
  \label{fig:G(T)}
\end{figure}

\emph{Finite T.}---As a benchmark for the reliability of our approach we first compare 
the linear conductance $G(T)$ with the NRG result. When plotted in units of
$T_K^*$, defined by $G(T=T_K^*)=1/2$, we see in Fig.~\ref{fig:G(T)} that our 
results in second and third order truncation fall almost on top of each other 
and agree very well with the NRG result \cite{costi_etal_94}.
The low- and high-temperature regimes can be studied analytically.
For $T\gg T_K^*$, we expand in $J_p(T)\ll 1$ and get
\begin{equation}
\label{eq:high_T}
G(T) \approx {3\pi^2 \over 4}J^2_p(T)\left(1-J_p(T)4\ln{2\pi\over e^{1+\gamma}}+O(J_p^2)\right) ,
\end{equation}
where $\gamma=0.577\dots$ is Euler's constant.
Besides the standard weak-coupling result, the second term is a third order correction 
which suppresses the conductance. For $T\ll T_K$, we expand in
$\bar{J}$ and obtain in third order truncation the Fermi-liquid result 
$G(T) \approx 1-c_T({T\over \bar{\Gamma}})^2$ with 
$c_T=\pi^4{\bar{J}^3(1+3\bar{J})\over (1-\bar{J})^3} + O(\bar{J}^5)$.
For the absolute value of the curvature w.r.t. $T_K^*$ we obtain in
third order truncation $c_T (T_K^*/\bar{\Gamma})^2\approx 4.8$, which is 
much better than the second order result but still not fully reliable 
due to the inaccuracy in the ratio $T_K^*/\bar{\Gamma}$ (the NRG result  
is about $10\%$ larger \cite{micklitz_etal_PRL06}).

\begin{figure}[t]
  \centering
  \includegraphics[width=70mm,clip=true]{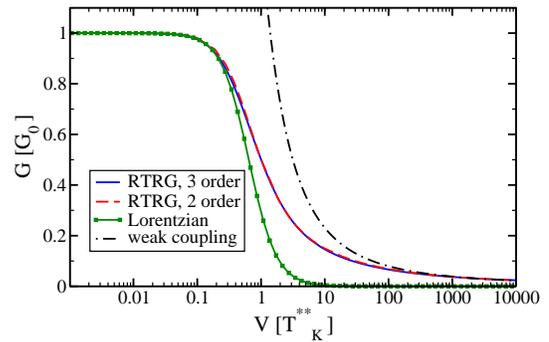}
  \caption{(Color online)
    $G(V)$ in second (red dashed line) and third (blue full line) order 
    truncation as function of $T/T_K^{**}$, defined by $G(V=T_K^{**})=1/2$,
    compared to Lorentzian fit (green squares) and high-$V$ result 
    (\ref{eq:high_V}) (dotted-dashed line).
    }
  \label{fig:G(V)}
\end{figure}

\emph{Finite V.}---Next we consider $T=0$ but finite $V$.
Figure~\ref{fig:G(V)} shows again that our second and third order truncation results for $G(V)$ 
agree very well with each other when plotted in units of $T_K^{**}$ defined by $G(V=T_K^{**})=1/2$.
For large $V\gg T_K^{**}$ we expand in $J_p(V)\ll 1$ and obtain
\begin{equation}
\label{eq:high_V}
G(V) \approx {3\pi^2 \over 4}J^2_p(V) + O(J_p^4) \quad.
\end{equation}
The first term is the standard weak-coupling result, but in addition
our third order analysis proves the absence of a term in $O(J_p^3)$.
For small $V\ll T_K^{**}$, we get $G(V) \approx 1-c_V({V\over \bar{\Gamma}})^2$ with 
$c_V/c_T=3/(2\pi^2)$. This coincides with the exact ratio known from Fermi-liquid theory 
\cite{oguri_JPSJ05,sela_malecki_PRB09}.
The result is fulfilled order by order by expanding in $\bar{J}$. 
The numerical solution does not show the exact ratio since inconsistent terms of 
$O(\bar{J}^5)$ are involved there, but in third order truncation the error is already below $10\%$.

\begin{figure}[t]
  \centering
  \includegraphics[width=75mm,clip=true]{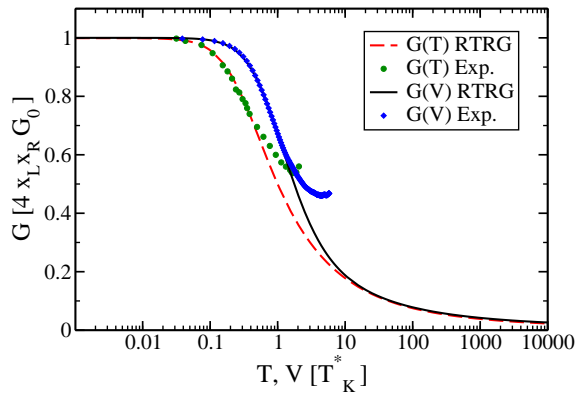}
  \caption{(Color online)
    Comparison  between the RTRG results and experimental data \cite{kretinin_etal_11} for $G(T)$ and $G(V)$. 
    Both  $T$ and $V$ are expressed in units  $T_K^*$ defined by $G(T=T_K^*)=1/2$. An asymmetry 
    factor $ 4 x_R x_L \approx 0.8$ is introduced to match the experimental $G(0)$ 
    with $ 4 x_R x_L$.}
  \label{fig:experiment}
\end{figure}

In Fig.~\ref{fig:G(V)} we show a comparison with a Lorentzian fit function 
$G(V)=1/[1+c_V(V/\bar{\Gamma})^2]$, as proposed in Ref.~\cite{konik_etal_01_02} 
from Bethe ansatz calculations. It turns out to be a good fit only for voltages
significantly below $T_K^{**}$, but becomes inconsistent with the weak-coupling solution when
applied in the whole crossover regime. The alternative trial function $G(V)=[1+(V/T_K^\prime(x))^2]^{-s}$
with $T_K^\prime(x)=T_K^{**}({1-b+bx^{s'}\over 2^{1/s}-1})^{1/2}$ and $x=V/T_K^{**}$, which is a good fit for 
$G(T)$ at $s=0.22$, $b=0$ and $T_K^{**}\rightarrow T_K^*$
\cite{costi_etal_94}, turns out to describe the crossover regime of $G(V)$ quite well at the
significantly different values $s=0.32$, $b=0.05$ and $s'=1.26$.
This shows that the universal line shapes of $G(V)$ and $G(T)$ are quite different and cannot
be rescaled on top of each other, as shown in Fig.~\ref{fig:experiment} when plotted in terms
of the same unit $T_K^*$. A numerical analysis gives $T_K^{**}/T_K^*\sim 1.8$ and we find
that, in third order truncation, $G(V)$ lies always above $G(T)$ for $V=T$, which can already be anticipated
from our analytical results at low and high energies. 
In Fig.~\ref{fig:experiment} we also show that recent experiments appear to be well captured
by the present theory. The experiments are performed on an 
InAs nanowire quantum dot \cite{kretinin_etal_11}, which can be represented as a single-level dot 
with the effective spin exchange $J\sim{2\Delta\over \pi U}\sim 0.28$.
The experimental data for $G(T)$ and $G(V)$ fit our results quite well in the regime $T,V\lesssim T_K^*$ 
using the {\it same} rescaling factor for the $V$ and $T$ dependence. Since $T_K^*\sim 0.1\,U$
in this experiment, universal features from spin fluctuations can only be 
expected below $T_K^*$, while above $T_K^*$ charge fluctuations will set in. 
As proposed in Ref.~\cite{kretinin_etal_preprint}, the universal relation
$G(V=T_K^*)\approx {2\over 3}$ allows for an elegant experimental determination of the
important energy scale $T_K^*$. The experimental conductance 
is not unitary due to uncontrolled asymmetric couplings to the two leads.
We have checked that asymmetric couplings
$J_{\alpha\alpha'}=2\sqrt{x_\alpha x_{\alpha'}}J$ with $0<x_\alpha<1$ and $x_L+x_R=1$
only give rise to the well-known asymmetry factor $4x_L x_R$ for the conductance in the weak 
and strong coupling regime \cite{hs_reininghaus_PRB09,sela_malecki_PRB09} and
change the normalized conductance $G(V)/(4x_L x_R)$ in the intermediate voltage regime by 
less than $1\%$ for the experimental value $4x_L x_R\sim 0.8$. Furthermore, we checked that 
the finite experimental temperature $T\sim 10\,mK \sim 0.025 \,T_K^*$ 
does not change $G(V)$ significantly.

\emph{Time evolution.}---At $T=V=0$ and for $t\gg 1/T_K$ we find from 
Eqs.~(\ref{eq:S(E)}) and (\ref{eq:rg}) that the spin dynamics is given by
$\langle \underline{S}\rangle(t)\sim {1\over t^g}e^{-\Gamma^*t}$.
The resolvent $1/[E+i\Gamma(E)]$ has a branch cut on the negative imaginary axis with a branching pole at $E=-i\Gamma^*$, which is identical to the fixed point $\tilde{J}=J^*$
of the RG flow and determines the rate of the exponential decay. 
We obtain the nearly constant ratio $\Gamma^*/T_K\approx 1.1$ independent of the truncation order.
To determine the exponent $g$ we expand around the fixed point and set $E=i\Lambda$ with 
$|\Lambda+\Gamma^*|\ll\Gamma^*$. We obtain
${dZ\over d\lambda}=4Z\tilde{J}^2/\lambda\approx\alpha Z/\lambda$ with $\alpha=4{J^*}^2$. 
This gives $Z\sim\lambda^\alpha$ and, by using $d_\Lambda\lambda=Z^{-1}\sim \lambda^{-\alpha}$, we
get $\lambda\sim (\Lambda+\Gamma^*)^{1/(1+\alpha)}$, 
leading to $g={\alpha\over 1+\alpha}={4{J^*}^2\over 1+4{J^*}^2}$
after the inverse Laplace transform.
However, the fixed point value $J^*$ depends on the truncation order. In second (third)
order truncation we get $J^*=1$ and $g={4\over 5}$ ($J^*={1\over 2}$ and $g={1\over 2})$.
As a result we obtain a universal exponential decay but the pre-exponential power law cannot be determined unambigiously.

In contrast, for $N$-channel Kondo models with $N>1$, where ${dZ\over d\lambda}=4NZ\tilde{J}^2/\lambda$ and
${d\tilde{J}\over d\lambda}=-2\tilde{J}^2(1-N\tilde{J})/\lambda$, 
the non-Fermi-liquid behavior at low energies 
\cite{cft,multi_channel} requires the fixed point to be at $E=0$. This gives rise to a pure power-law decay 
for the spin dynamics in the long-time limit. The exponent follows from $\alpha=4N{J^*}^2$ and
$J^*={1\over N}$ to be $g={4N{J^*}^2\over 1+4N{J^*}^2}={4\over 4+N}$.
However, the precise value of the exponent is only reliable for $N\gg 1$,
where the $\beta$-function can be systematically truncated \cite{mitra_rosch_PRL11}. 

\emph{Conclusions.}---We have derived an approximate form for the
universal line shape of the zero-bias anomaly for the isotropic Kondo model,
which successfully describes the recent experiments
\cite{kretinin_etal_11,kretinin_etal_preprint}. The reliability of our
approach has been checked by comparing different truncation orders and
by reproducing the correct asymptotic values at large and small voltages. 
In the long-time limit we found a universal exponential decay 
for the spin dynamics in the $1$-channel case and a power-law decay for
multichannel models. Our approximate results could serve as a reference for future more
refined numerical and analytical approaches.
In particular, the precise value of the exponential decay rate and the form of the
pre-exponential modulation is an open issue. The new 
formulation of the RTRG method using the Laplace variable as the flow parameter is a promising 
analytical tool for solving other nonequilibrium problems of quantum impurity 
physics and dissipative quantum mechanics in the strong coupling regime. 

We acknowledge valuable discussions with S. Andergassen, F. Anders, N. Andrei,
H. Capellmann, T. Costi, J. von Delft, R. Konik, A. Kretinin, V. Meden, F. Reininghaus,
P. Schmitteckert, E. Sela, D. Schuricht, A. Weichselbaum, and G. Zarand, and 
financial support from DFG-FG 723.

\end{document}


\title{Nonequilibrium Kondo model: Crossover from weak to strong coupling\\
Supplementary Material}

\author{Mikhail Pletyukhov}
\affiliation{Institut f\"ur Theorie der Statistischen Physik, RWTH Aachen, 
52056 Aachen, Germany and JARA - Fundamentals of Future Information Technology}
\author{Herbert Schoeller}
\affiliation{Institut f\"ur Theorie der Statistischen Physik, RWTH Aachen, 
52056 Aachen, Germany and JARA - Fundamentals of Future Information Technology}
\maketitle
The following appendices comprise the supplementary material to Ref.~\onlinecite{pletyukhov_hs}.

\section{The $E$-flow scheme of RTRG}
\label{sec:E_flow}
In Ref.~\onlinecite{hs_review_EPJ09} a real-time renormalization group
method (RTRG) has been developed in Liouville space, which can be used to evaluate the effective
Liouvillian $L(E)$ of a local quantum system coupled to several
fermionic reservoirs. Within this method, a Matsubara cutoff was defined by cutting off the 
Matsubara poles of the Fermi functions. A systematic weak-coupling
expansion including cutoff scales from relaxation and decoherence 
rates was developed, which was used to calculate the nonequilibrium
properties of the Kondo model \cite{hs_reininghaus_PRB09} and the interacting resonant
level model (IRLM) \cite{IRLM_RTRG} in the weak coupling regime.  
In the strong coupling regime and in the presence of spin, orbital or potential fluctuations, a problem 
within the RTRG method is the generation of spurious terms linear in the physical band width 
$2D$ of the reservoirs, originating from the fact that the cutoff used in the RG is not identical 
to the physical band width. In particular, it turns out that the RG-flow is unstable against a small
variation of these terms. In the weak-coupling regime it has been shown
\cite{hs_review_EPJ09,hs_reininghaus_PRB09} that the problem
can be resolved by means of an appropriate subtraction scheme. The same was demonstrated for the
strong coupling regime of the IRLM \cite{IRLM_RTRG}. However, in the generic
strong coupling case it is not clear whether the subtraction scheme can be uniquely defined.

Here we propose an alternative flow scheme where this problem is avoided by
solving the RG-equations starting at low energies. It uses the Laplace variable $E$
as flow parameter and for this reason is called $E$-flow scheme in the following. Another advantage
of the $E$-flow scheme consists in the fact that the Laplace variable is a physical energy scale from
which the time evolution can be directly evaluated. The effective Liouvillian and the 
vertices occurring in the RG equations are the full physical ones. Furthermore, the 
method allows the consideration of all diagrams up to a certain order without 
adhering to approximations typically used in the weak-coupling regime.

We show here the derivation of the $E$-flow RG equations in third order truncation for the case
where only fermionic $2$-point vertices $G_{12}(E,\bar{\omega}_1,\bar{\omega}_2)$ 
are present and  describe either spin, orbital or potential fluctuations 
(higher-order vertices are only generated in fourth order). Moreover, we assume
that the bare vertices have no explicit frequency-dependence and we take a flat d.o.s. in
the reservoirs with band width $2D$. The index $1\equiv \eta\alpha\sigma$ characterizes the reservoir field
operator: $\eta=\pm$ distinguishes creation/annihilation operators, $\alpha$ is
the reservoir index, and $\sigma$ is the channel index (i.e. the spin for the Kondo model). The frequency arguments
$\bar{\omega}_i=\eta_i\omega_i$ contain the reservoir energy $\omega_i$ measured relative to the 
chemical potential $\mu_\alpha$ of reservoir $\alpha$. The Keldysh index $p$ can also be included in 
the index $1$ but is disregarded here since, as will be shown below, only the average over the 
Keldysh indices remains in the final RG equations. Including the Keldysh indices 
the contraction between two vertices is given by
$\gamma_{11'}^{pp'}(\omega,\omega')=\delta_{1\bar{1}'}\delta(\bar{\omega}+\bar{\omega}')
\gamma^{p'}(\bar{\omega})$, where 
$\gamma^{p'}(\bar{\omega})=p'f(p'\bar{\omega}){D^2\over D^2+\bar{\omega}^2}$, 
$1=\eta\alpha\sigma$, $\bar{1}'=-\eta'\alpha'\sigma'$, and $f(\omega)={1\over e^{\omega/T}+1}$ 
denotes the Fermi function at temperature $T$.

To achieve convergence of the frequency integrals in the scaling limit $D\rightarrow\infty$ one needs 
equations for the second derivative of $L(E)$ and
the first derivative of $G_{12}(E,\bar{\omega}_1,\bar{\omega}_2)$ w.r.t. $E$. Using the diagrammatic representation
described in Ref.~\onlinecite{hs_review_EPJ09}, the $E$-derivative can only act on the
propagators $\Pi(E)={1\over E-L(E)}$ occurring between the bare vertices (we always resum
all self-energy insertions such that the full effective Liouvillian appears in the
propagator). Resumming all diagrams right and left to the propagator in which the $E$-derivative
is taken, we obtain the following two equations up to $O(G^3)$
\begin{align}
\label{eq:L_rg_1}
{1\over 2}{\partial^2\over \partial E^2}L(E)\,&=\,
{1\over 2}{1\over 2}
\begin{picture}(10,10)
\put(5,-7){\includegraphics[height=0.7cm]{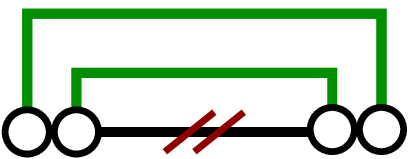}}
\end{picture}
\hspace{2cm}+
\begin{picture}(10,10)
\put(10,-7){\includegraphics[height=0.7cm]{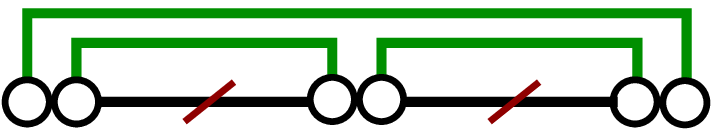}}
\end{picture}
\hspace{4.5cm}+\quad O(G^4) \quad,
\\ \nonumber \\ \nonumber
{\partial\over \partial E}G_{12}(E,\bar{\omega}_1,\bar{\omega}_2)\, & =\,
\left(
\begin{picture}(10,10)
\put(5,-12){\includegraphics[height=0.8cm]{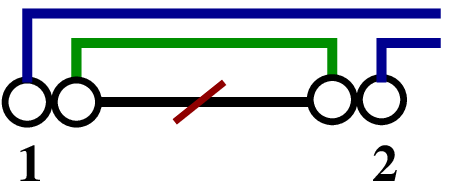}}
\end{picture}
\hspace{2cm}-(1\leftrightarrow 2)\right)
\quad + \quad 
{1\over 2}
\begin{picture}(10,10)
\put(5,-12){\includegraphics[height=1cm]{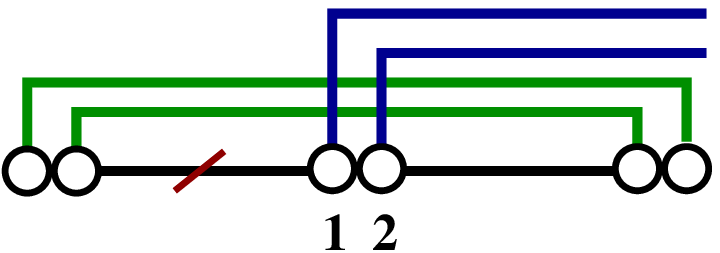}}
\end{picture}
\hspace{2.8cm}
\quad + \quad 
{1\over 2}
\begin{picture}(40,10)
\put(5,-12){\includegraphics[height=1cm]{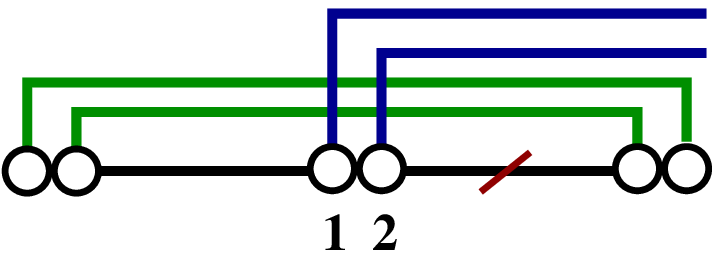}}
\end{picture}
\\ \nonumber \\ 
\label{eq:G_rg_1}
& \quad + \quad
\left(
\begin{picture}(10,10)
\put(1,-12){\includegraphics[height=1cm]{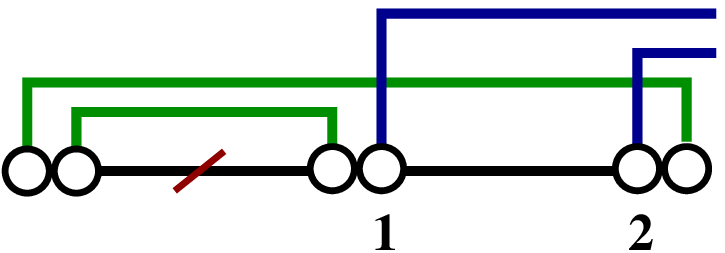}}
\end{picture}
\hspace{3cm}+\quad
\begin{picture}(10,10)
\put(1,-12){\includegraphics[height=1cm]{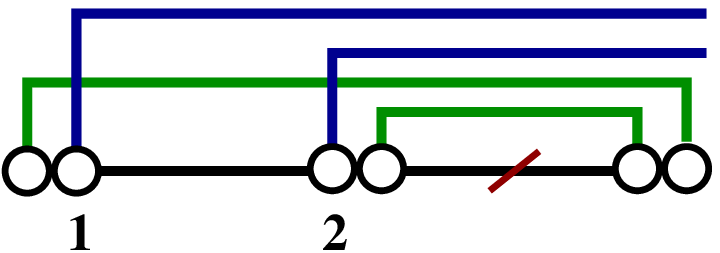}}
\end{picture}
\hspace{3cm}-\quad(1\leftrightarrow 2)\right)
\quad + \quad 
O(G^4) \quad.
\end{align}
Here the slash indicates the $E$-derivative ${\partial\over\partial E}$ and a
double-circle represents the full effective $2$-point vertex (containing all
connected diagrams with two free reservoir lines). Symmetry factors ${1\over n!}$ 
arising either from the diagrammatic rules (when two vertices are connected by $n$ equivalent
lines) or from the $E$-derivatives ${1\over n!}\partial^n_E$ are explicitly quoted. Since the propagators 
$\Pi_{1\dots n}=\Pi(E_{1\dots n}+\bar{\omega}_{1\dots n})$
between the vertices contain the variables 
$E_{1\dots n}=E+\bar{\mu}_{1\dots n}=E+\bar{\mu}_1+\dots \bar{\mu}_n$
and $\bar{\omega}_{1\dots n}=\bar{\omega}_1+\dots \bar{\omega}_n$ with
$\bar{\mu}_i=\eta_i\mu_{\alpha_i}$ and $\bar{\omega}_i=\eta_i\omega_i$, the
$E$-derivative can be written as frequency-derivative ${\partial\over\partial\bar{\omega}_i}$
and we can apply partial integration to calculate the frequency integrals.
E.g., for the first term on the r.h.s. of (\ref{eq:G_rg_1}) we obtain (we have
permuted the two indices of the vertices by using antisymmetry 
$G_{12}(E,\bar{\omega}_1,\bar{\omega}_2)=-G_{21}(E,\bar{\omega}_2,\bar{\omega}_1)$)
\begin{align}
\nonumber
\begin{picture}(10,10)
\put(5,-12){\includegraphics[height=1cm]{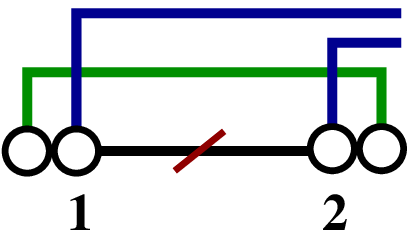}}
\end{picture}
\hspace{2cm}\,&=\,\quad-
\begin{picture}(10,10)
\put(5,-12){\includegraphics[height=1cm]{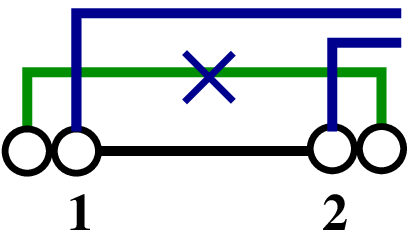}}
\end{picture}
\hspace{2cm}-\quad 
\begin{picture}(10,10)
\put(5,-12){\includegraphics[height=1cm]{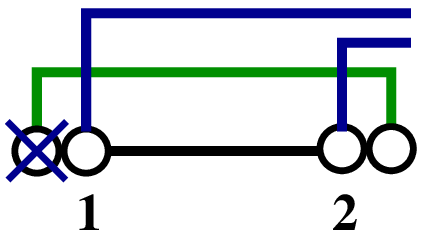}}
\end{picture}
\hspace{2cm}-\quad 
\begin{picture}(10,10)
\put(5,-12){\includegraphics[height=1cm]{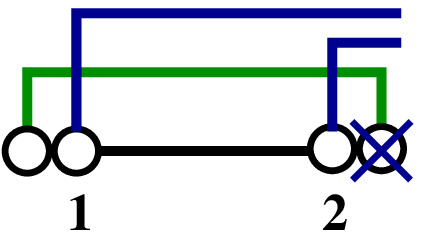}}
\end{picture}
\\ \nonumber\\ 
\label{eq:G_partial}
&=\,\quad-
\begin{picture}(10,10)
\put(5,-12){\includegraphics[height=1cm]{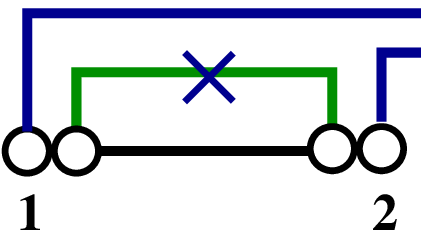}}
\end{picture}
\hspace{2cm}-\quad 
\begin{picture}(10,10)
\put(5,-12){\includegraphics[height=1cm]{G_Eder_4.eps}}
\end{picture}
\hspace{3cm}-\,\, 
\begin{picture}(10,10)
\put(5,-12){\includegraphics[height=1cm]{G_Eder_5.eps}}
\end{picture}
\hspace{3cm}+\quad O(G^4) \quad.
\end{align}
Here the cross indicates the frequency derivative w.r.t. $\bar{\omega}$ of either the corresponding
reservoir contraction or the frequency argument of the vertices. The first line
is exact and follows from partial integration. For the derivation of the
second line we have used 
\begin{equation}
\label{eq:G_cross}
\begin{picture}(10,10)
\put(5,-15){\includegraphics[height=1cm]{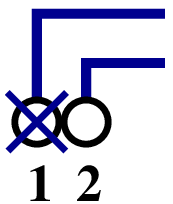}}
\end{picture}
\hspace{1cm}  = 
\quad 
\begin{picture}(10,10)
\put(5,-15){\includegraphics[height=1cm]{G_Eder_1.eps}}
\end{picture}
\hspace{2.5cm}+\quad O(G^3) 
\hspace{1cm},\hspace{1cm}
\begin{picture}(10,10)
\put(5,-15){\includegraphics[height=1cm]{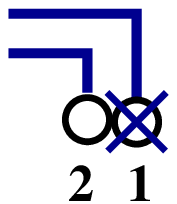}}
\end{picture}
\hspace{1cm}  = 
\quad 
\begin{picture}(10,10)
\put(5,-15){\includegraphics[height=1cm]{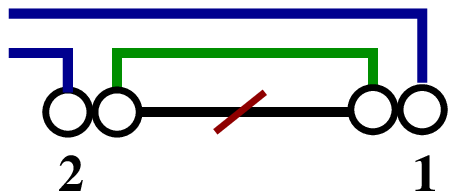}}
\end{picture}
\hspace{2.5cm}+\quad O(G^3) \quad,
\end{equation}
which follows analogously to (\ref{eq:G_rg_1}) by using the fact that, in the
original diagrammatic series, a frequency associated with an external line occurs only in those 
propagators which lie below the external line but not in other propagators and not
in the bare vertices. Note that these relations
can only be applied if it is specified whether the external line involving
the frequency derivative is directed either towards the left or towards the right. 

Analogously, we can treat the first term on the r.h.s. of (\ref{eq:L_rg_1}) by two 
partial integrations and using the fact that 
${\partial^2\over\partial\bar{\omega}_1\partial\bar{\omega}_2}
G_{12}(E,\bar{\omega}_1,\bar{\omega}_2)\sim O(G^3)$.
We obtain
\begin{align}
\nonumber
\begin{picture}(10,10)
\put(5,-7){\includegraphics[height=0.7cm]{L_Eder_1.eps}}
\end{picture}
\hspace{2cm}\,&=\,
\begin{picture}(10,10)
\put(5,-7){\includegraphics[height=0.8cm]{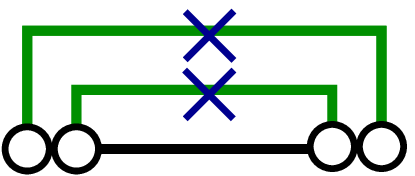}}
\end{picture}
\hspace{2cm}+\quad 2
\begin{picture}(10,10)
\put(5,-7){\includegraphics[height=0.75cm]{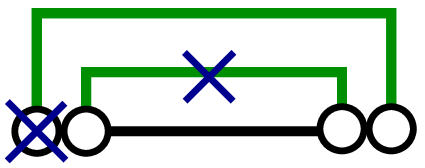}}
\end{picture}
\hspace{2cm}+\quad 2
\begin{picture}(10,10)
\put(5,-7){\includegraphics[height=0.75cm]{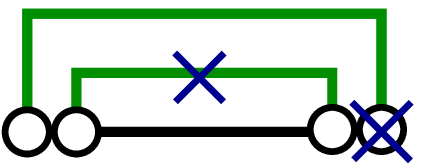}}
\end{picture}
\hspace{2cm}+\quad O(G^4)
\\ \nonumber \\
\label{eq:L_partial} 
&=\,
\begin{picture}(10,10)
\put(5,-7){\includegraphics[height=0.8cm]{L_cross_1.eps}}
\end{picture}
\hspace{2cm}-\quad 4
\begin{picture}(10,10)
\put(5,-7){\includegraphics[height=0.75cm]{L_Eder_2.eps}}
\end{picture}
\hspace{4.5cm}+\quad O(G^4) \quad,
\end{align}
where, in the second step, we have again used (\ref{eq:G_cross}).

Inserting (\ref{eq:L_partial}) and (\ref{eq:G_partial}) in
(\ref{eq:L_rg_1}) and (\ref{eq:G_rg_1}), respectively, we see
that many diagrams in $O(G^3)$ cancel each other and we cast
the final RG-equations to a very compact and generic
form 
\begin{align}
\label{eq:L_rg_2}
{\partial^2\over \partial E^2}L(E)\quad &= \quad
{1\over 2}
\begin{picture}(10,10)
\put(5,-7){\includegraphics[height=0.7cm]{L_cross_1.eps}}
\end{picture}
\hspace{2cm}+\quad O(G^4),
\\ \nonumber \\
\label{eq:G_rg_2}
{\partial\over \partial E}G_{12}(E,\bar{\omega}_1,\bar{\omega}_2)\quad &= \quad
-\left(
\begin{picture}(10,10)
\put(5,-12){\includegraphics[height=1cm]{G_cross_1v.eps}}
\end{picture}
\hspace{2cm}-(1\leftrightarrow 2)\right)
\quad - \quad 
{1\over 2}
\begin{picture}(10,10)
\put(5,-12){\includegraphics[height=1.2cm]{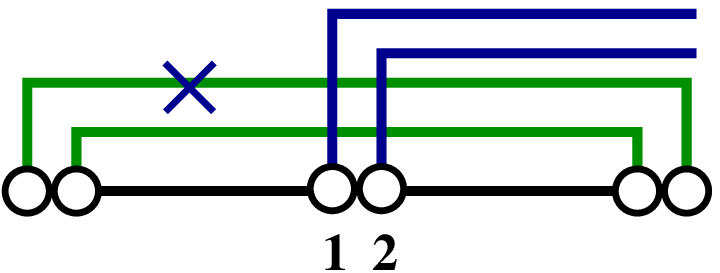}}
\end{picture}
\hspace{3.5cm}+\quad O(G^4).
\end{align}

The RG equations are constructed in such a way that all frequency integrals on the r.h.s. 
of the RG equations are well-defined in the 
limit $D\rightarrow\infty$ even when the frequency dependence of the vertices is neglected.
Performing this limit, each reservoir contraction contains
$\gamma^{p'}(\bar{\omega})=p'f(p'\bar{\omega})$.
As a consequence, the frequency integral for a contraction where the derivative is taken
(diagrammatically indicated by the cross) involves 
${\partial\over\partial\bar{\omega}}\gamma^{p'}(\bar{\omega})=f'(\bar{\omega})$, which is
independent of the Keldysh indices. The frequency integral of the contraction without a
derivative in the second term on the r.h.s. of (\ref{eq:G_rg_2}) involves
$\gamma^{p'}(\bar{\omega})=f^a(\bar{\omega})+{p'\over 2}$, where $f^a(\omega)=f(\omega)-{1\over 2}$
is the antisymmetric part of the Fermi function. Since two propagators are involved in this
frequency integral, the integration contour can be closed in the upper half of the complex plane
and the symmetric part ${p'\over 2}$ vanishes identically since the resolvents and the vertices
are analytic functions in the upper half plane \cite{hs_review_EPJ09}. Therefore, an explicit dependency on the
Keldysh indices disappears, and it is sufficient to consider the vertices 
$G_{12}(E,\bar{\omega}_1,\bar{\omega}_2)=\sum_{p_1 p_2}G^{p_1 p_2}_{12}(E,\bar{\omega}_1,\bar{\omega}_2)$
averaged over the Keldysh indices. We note that the symmetric part of the Fermi function
does not influence the RG equations but enters into the initial conditions at high energies 
\cite{hs_review_EPJ09}.

To evaluate the frequency integrals on the r.h.s. of the RG equations (\ref{eq:L_rg_2}) and 
(\ref{eq:G_rg_2}) explicitly, we need a consistent approximation for the frequency 
dependence of the vertices and the Liouvillian. For the vertices, this
can be achieved by formally integrating the first two terms on the r.h.s. of (\ref{eq:G_rg_2})
yielding
\begin{equation}
\label{eq:G_omega}
\begin{picture}(10,10)
\put(5,-10){\includegraphics[height=0.8cm]{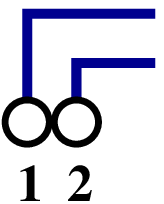}}
\end{picture}
\hspace{1cm}=\quad
\begin{picture}(10,10)
\put(5,-10){\includegraphics[height=0.8cm]{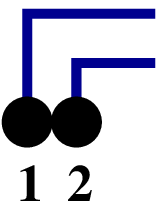}}
\end{picture}
\hspace{1cm}+\quad
\begin{picture}(10,10)
\put(5,-13){\includegraphics[height=1cm]{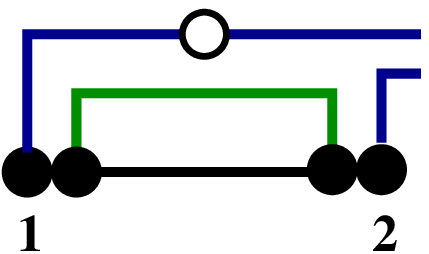}}
\end{picture}
\hspace{2cm}- \quad
\begin{picture}(10,10)
\put(5,-13){\includegraphics[height=1cm]{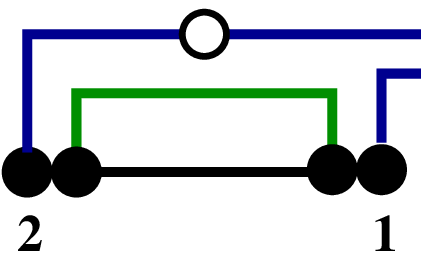}}
\end{picture}
\hspace{2cm} + \quad 
O(G^3) \quad,
\end{equation}
where $G_{12}(E)=G_{12}(E,0,0)$ is indicated by filled double dots. The open circle at a 
contraction indicates that the propagator corresponding to the vertical cut at the position of 
that circle has to be replaced by the difference 
$\Pi(E_{1\dots n}+\bar{\omega}_{1\dots n}+\bar{\omega})-\Pi(E_{1\dots n}+\bar{\omega}_{1\dots n})$
of the two propagators with and without the frequency $\bar{\omega}$ corresponding to
the contraction with the circle. The identity (\ref{eq:G_omega}) refers to the case where
the two external lines are directed to the right, similar equations can be written down for
the other cases. We note that the sign factor in the second term on the r.h.s. accounts explicitly
for the crossing of the two external lines relative to the first term. Therefore, when using 
(\ref{eq:G_omega}) in a certain diagram the sign factor must not be written explicitly since
it is automatically accounted for in the diagrammatic rules.
Inserting (\ref{eq:G_omega}) in (\ref{eq:L_rg_2}) and (\ref{eq:G_rg_2}) we obtain
\begin{align}
\label{eq:L_rg_3}
{\partial^2\over \partial E^2}L(E)\quad &= \quad
{1\over 2}
\begin{picture}(10,10)
\put(5,-7){\includegraphics[height=0.7cm]{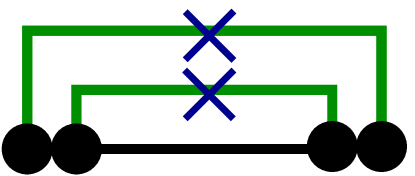}}
\end{picture}
\hspace{2cm}+\quad
\begin{picture}(10,10)
\put(5,-7){\includegraphics[height=0.7cm]{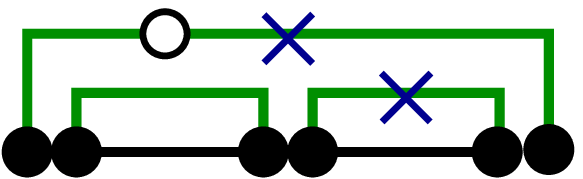}}
\end{picture}
\hspace{3cm}+\quad
\begin{picture}(10,10)
\put(5,-7){\includegraphics[height=0.7cm]{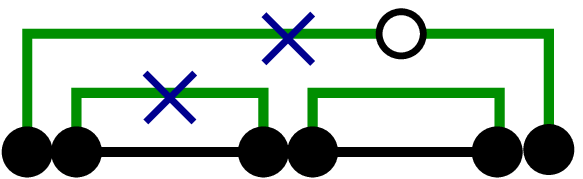}}
\end{picture}
\hspace{3cm}+\quad O(G^4),
\\ \nonumber \\ \nonumber
{\partial\over \partial E}G_{12}(E)\quad &= \quad
-\left(
\begin{picture}(10,10)
\put(5,-12){\includegraphics[height=1cm]{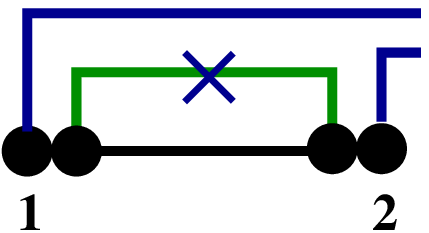}}
\end{picture}
\hspace{2cm}+
\begin{picture}(10,10)
\put(5,-12){\includegraphics[height=1cm]{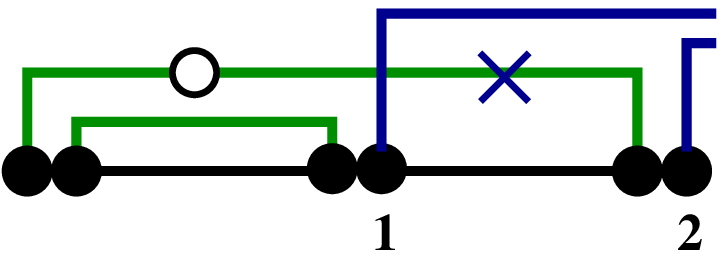}}
\end{picture}
\hspace{3cm}+
\begin{picture}(10,10)
\put(5,-12){\includegraphics[height=1cm]{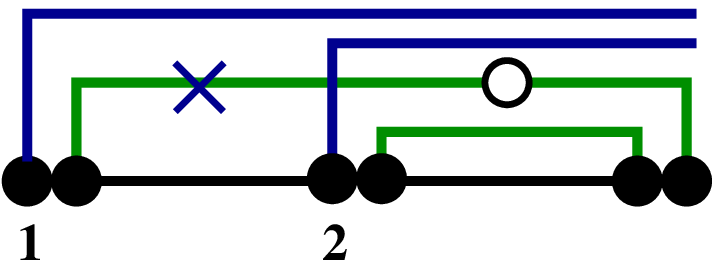}}
\end{picture}
\hspace{3cm}-(1\leftrightarrow 2)\right)
\\ \nonumber \\
\label{eq:G_rg_3}
& \hspace{2cm}-\quad {1\over 2}
\begin{picture}(10,10)
\put(5,-12){\includegraphics[height=1.2cm]{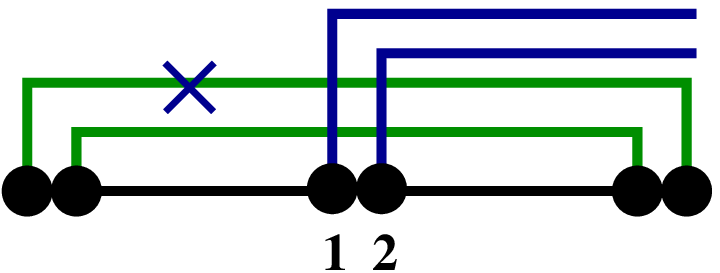}}
\end{picture}
\hspace{3.5cm}+\quad O(G^4) \quad.
\end{align}
Finally, to calculate the frequency integrals in these equations, one needs a consistent
approximation for the propagator $\Pi(E+\bar{\omega})$, where $\bar{\omega}$ is an integration variable.
Using ${\partial^2\over\partial E^2}L(E)\sim O({1\over E}G^2)$ cf. (\ref{eq:L_rg_3}), we approximate
$L(E+\bar{\omega})\approx L(E)+{\partial\over\partial E}L(E)\bar{\omega}$.
This leads to the following approximation for the propagator
\begin{equation}
\label{eq:prop}
\Pi(E+\bar{\omega})\,\approx\,{1\over \bar{\omega} + \chi(E)}\,Z(E) \quad,
\end{equation}
where $Z(E)={1\over 1-{\partial\over \partial E}L(E)}$ is the $Z$-factor operator and
$\chi(E)=Z(E)(E-L(E))$ is an operator defining the distance to resonance positions. We note that
the eigenvalue zero of the Liouvillian $L(E)$ (or of $Z(E)L(E)$) can be omitted in (\ref{eq:prop}) since 
the projector $P_0$ on the corresponding eigenvector fulfils $P_0 L(E)=P_0 G_{12}=0$ and 
$P_0 Z(E)=1$ \cite{hs_review_EPJ09}. Using (\ref{eq:prop})
in (\ref{eq:L_rg_3}) and (\ref{eq:G_rg_3}), all frequency integrals can be
calculated analytically, which will be done explicitly for the Kondo model in the
next section.

Equations (\ref{eq:L_rg_3}) and (\ref{eq:G_rg_3}) along with (\ref{eq:prop}) are the final 
RG equations for a determination of  the effective Liouvillian $L(E)$ from which the reduced density
matrix of the local system in Laplace space follows via $\rho(E)={i\over E-L(E)}\rho_0$,
where $\rho_0=\rho(t=0)$ is the initial condition. To find the average of the current 
$I_\alpha=-i\text{Tr}\Sigma_\alpha(E)\rho(E)$ flowing into reservoir $\alpha$, one needs the
current kernel $\Sigma_\alpha(E)$ in Laplace space. As shown in Ref.~\onlinecite{hs_review_EPJ09}
it can be determined analogously to $L(E)$ by replacing the first vertex from the left in all diagrams of
(\ref{eq:L_rg_3}) and (\ref{eq:G_rg_3}) by the current vertex $I^\alpha_{12}(E)$. To calculate the
differential conductance one needs the variation $\delta I_\alpha$ for an infinitesimal variation
$\delta\mu_\alpha$ of the chemical potentials of the reservoirs. Within the $E$-flow scheme the
RG equation for ${\partial\over \partial E}\delta L(E)$ (or, equivalently, for 
${\partial\over\partial E}\delta \Sigma_\alpha(E)$ by
replacing the first vertex by the current vertex) can be  straightforwardly established by applying
the variation to the original diagrammatic series and using
$\delta\Pi_{1\dots n}=\delta\bar{\mu}_{1\dots n}{\partial\over \partial E}\Pi_{1\dots n}+\Pi_{1\dots n}
\delta L_{1\dots n}\Pi_{1\dots n}$, where 
$\delta L_{1\dots n}=(\delta L)(E_{1\dots n}+\bar{\omega}_{1\dots n})$. 
After resummation we obtain by analogy with (\ref{eq:L_rg_1}) up to $O(G^3)$
\begin{align}
\nonumber
{\partial\over \partial E}\,\delta L(E)\quad & = 
\quad {1\over 2}\,\delta\bar{\mu}_{12}
\begin{picture}(10,10)
\put(5,-15){\includegraphics[height=1cm]{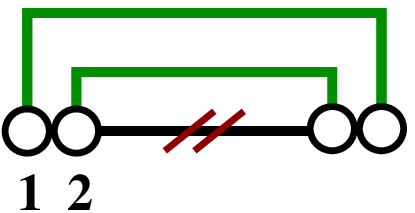}}
\end{picture}
\hspace{2cm}+
\quad {1\over 2}
\begin{picture}(10,10)
\put(5,-10){\includegraphics[height=0.8cm]{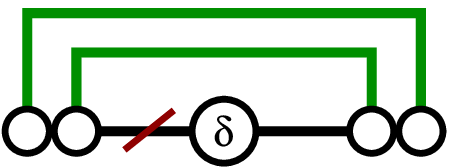}}
\end{picture}
\hspace{2.5cm}+
\quad {1\over 2}
\begin{picture}(10,10)
\put(5,-10){\includegraphics[height=0.8cm]{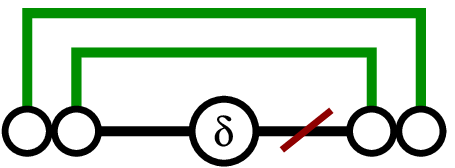}}
\end{picture}
\\ \nonumber \\
\label{eq:L_rg_V1}
&\hspace{2cm}+
\quad (\delta\bar{\mu}_{12}+\delta\bar{\mu}_{13})
\begin{picture}(10,10)
\put(5,-12){\includegraphics[height=1cm]{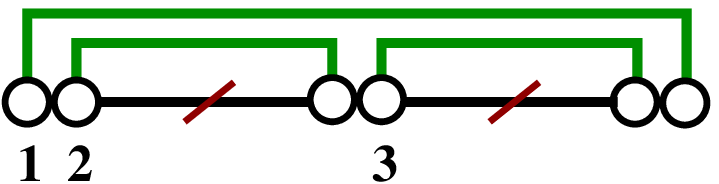}}
\end{picture}
\hspace{4.5cm}+\quad O(G^4) \quad,
\end{align}
where $\delta L\sim O(\delta\mu G)$ is represented by 
$\begin{picture}(10,10)
\put(1,-4){\includegraphics[height=0.4cm]{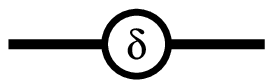}}
\end{picture}\hspace{1.3cm}$ and can be calculated from the first (lowest order)
term in the r.h.s. of (\ref{eq:L_rg_V1}) and subsequently inserted in the second and third term
in the r.h.s. of (\ref{eq:L_rg_V1}). Applying two partial integrations to the
first term on the r.h.s. of (\ref{eq:L_rg_V1}) by analogy with (\ref{eq:L_partial}),
we obtain
\begin{equation}
\label{eq:L_rg_V2}
{\partial\over \partial E}\,\delta L(E)\quad  = 
\quad {1\over 2}\,\delta\bar{\mu}_{12}
\begin{picture}(10,10)
\put(5,-15){\includegraphics[height=1cm]{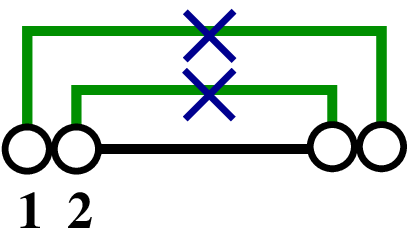}}
\end{picture}
\hspace{2cm}-
\quad {1\over 2}
\begin{picture}(10,10)
\put(5,-10){\includegraphics[height=1cm]{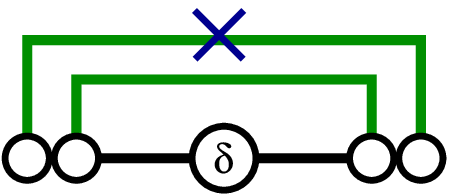}}
\end{picture}
\hspace{2.5cm}+\quad O(G^4) \quad.
\end{equation}
Using (\ref{eq:G_omega}) we get the final RG equation for the variation of the
kernel
\begin{align}
\nonumber
{\partial\over \partial E}\,\delta L(E)\quad & = 
\quad {1\over 2}\,\delta\bar{\mu}_{12}
\begin{picture}(10,10)
\put(5,-15){\includegraphics[height=1cm]{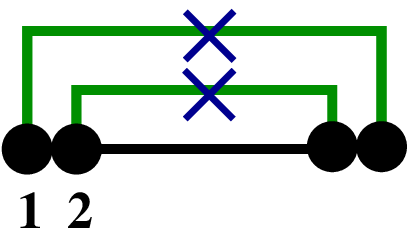}}
\end{picture}
\hspace{2cm}-
\quad {1\over 2}
\begin{picture}(10,10)
\put(5,-10){\includegraphics[height=1cm]{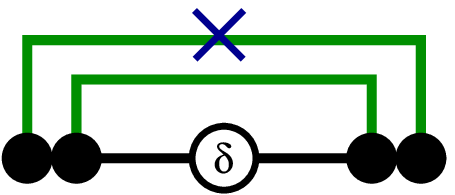}}
\end{picture}
\\ \nonumber \\
\label{eq:L_rg_V3}
&\hspace{1cm}+\quad \delta\bar{\mu}_{13}
\begin{picture}(10,10)
\put(5,-12){\includegraphics[height=1cm]{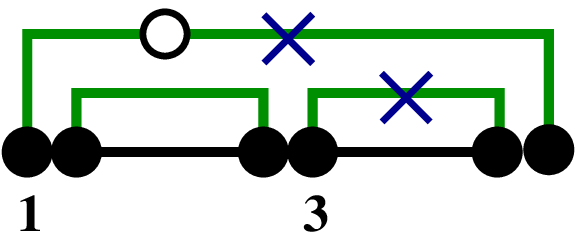}}
\end{picture}
\hspace{3cm}+\quad \delta\bar{\mu}_{12}
\begin{picture}(10,10)
\put(5,-12){\includegraphics[height=1cm]{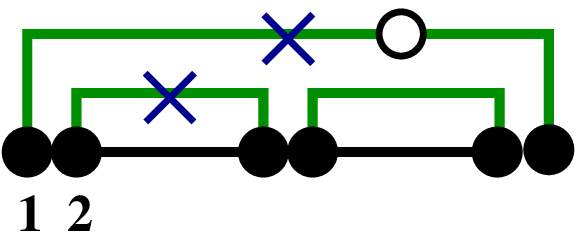}}
\end{picture}
\hspace{3cm}+\quad O(G^4) \quad.
\end{align}
We note that at zero temperature all diagrams in (\ref{eq:L_rg_3}), (\ref{eq:G_rg_3}) 
and (\ref{eq:L_rg_V3}) containing a contraction with a circle and a cross vanish,
since the cross restricts the frequency to zero value.

\section{RG equations for the Kondo model}
\label{sec:RGKondo}
We now apply the RG-equations (\ref{eq:L_rg_3}), (\ref{eq:G_rg_3}) and (\ref{eq:L_rg_V3}) to 
the isotropic spin-${1\over 2}$ Kondo model at zero magnetic field. 
Using spin conservation and rotational invariance it has been shown in Ref.~\onlinecite{hs_review_EPJ09}
how the kernels and the vertices can be decomposed into convenient basis superoperators. 
One obtains $L(E)=-i\Gamma(E)L^a$, $\Sigma_\gamma(E)=i\Gamma_\gamma(E)L^b$,  
$G_{+\alpha\sigma,-\alpha'\sigma'}(E)=\sum_{\chi=a,2,3}G^{\chi}_{\alpha\alpha'}(E)\hat{L}^\chi_{\sigma\sigma'}$,
and $I^\gamma_{+\alpha\sigma,-\alpha'\sigma'}(E)=
\sum_{\chi=b,1}I^{\gamma,\chi}_{\alpha\alpha'}(E)\hat{L}^\chi_{\sigma\sigma'}$,
where $\hat{L}^{a,b}_{\sigma\sigma'}=L^{a,b}\delta_{\sigma\sigma'}$ and 
$\hat{L}^{i}_{\sigma\sigma'}=\underline{L}^{i}\underline{\sigma}_{\sigma\sigma'}$ for $i=1,2,3$
($\underline{\sigma}$ are the Pauli matrices).
The superoperators $L^{a,b}$ and $\underline{L}^{1,2,3}$ can be expressed in terms of the 
spin superoperators $\underline{L}^+=\underline{S}\cdot$ and $\underline{L}^-=\cdot(-\underline{S})$ via
$L^a={3\over 4}+\underline{L}^+\underline{L}^-$, $L^b={1\over 4}-\underline{L}^+\underline{L}^-$,
$\underline{L}^{1,3}={1\over 2}(\underline{L}^+ - \underline{L}^-\mp 2i \underline{L}^+ \wedge \underline{L}^-)$,
and $\underline{L}^{2}=-{1\over 2}(\underline{L}^+ + \underline{L}^-)$.
The closed algebra of these superoperators needed to evaluate the RG-equations can be found
in Ref.~\onlinecite{hs_review_EPJ09}. The algebra is applied after the summations over the 
creation/annihilation and spin indices $\eta$ and $\sigma$ have been performed. Furthermore,
by using (\ref{eq:prop}), the intermediate propagators are replaced by the c-numbers 
$\Pi(E+\bar{\omega})\approx{Z(E)\over \bar{\omega}+\chi(E)}$, where $Z(E)={1\over 1+i{d\Gamma\over dE}(E)}$
and $\chi(E)=Z(E)(E+i\Gamma(E))$. This follows from the fact that the zero eigenvalue of
the Liouvillian can be omitted, and the three nonzero eigenvalues are degenerate and given 
by $-i\Gamma(E)$. 

A detailed analysis shows that the vertex component $I^{\gamma,b}$ is generated but does not influence 
the current kernel. Furthermore, with $G^2\sim J$, it turns out that the vertex component $G^a$ is 
generated in $O(J^2)$ and 
influences the Liouvillian only in $O(J^4)$. Therefore it can be omitted as well and we are left 
with a set of closed equations for the vertex components $G^2$, $G^3$ and $I^{\gamma,1}$, together with
the rates $\Gamma$ and $\Gamma_\gamma$. For applied voltages the energy arguments are shifted by
multiples of the chemical potentials of the reservoirs and all components couple to each other. 
After the summations over the indices $\eta$ and $\sigma$ have been performed, we use the
shorthand notation $1\equiv \alpha$ for the reservoir index. Furthermore, we define
$\mu_{12}=\mu_{\alpha_1}-\mu_{\alpha_2}$, $E_{12}=E+\mu_{12}$ and $E_{1234}=E+\mu_{12}+\mu_{34}$. 
For convenience we define $J_{12}(E)=-G^2_{12}(E)$, $K_{12}(E)=-i{2\over \pi}G^3_{12}(E_{21})$ and
$I^\gamma_{12}(E)=-4 I^{\gamma,1}_{12}(E)$. From hermiticity of the original Hamiltonian, we 
obtain the symmetry properties
$J_{12}(E)^*=J_{21}(-E^*)$, $K_{12}(E)^*=K_{21}(-E^*)$ and $I^\gamma_{12}(E)^*=-I^\gamma_{21}(-E^*)$
for the vertex components, and $\Gamma(E)^*=\Gamma(-E^*)$ and $\Gamma_\gamma(E)^*=\Gamma_\gamma(-E^*)$
for the rates \cite{hs_review_EPJ09}. The initial conditions at high energies arising from 
the symmetric part of the Fermi function are calculated in Ref.~\onlinecite{hs_review_EPJ09} as
\begin{align}
\label{eq:initial}
Z\,&=\,1\quad, & K_{\alpha\alpha'}\,&=\, \sum_{\alpha''}J_{\alpha\alpha''}J_{\alpha''\alpha'}\quad, & 
I^\gamma_{\alpha\alpha'}\,&=\,(\delta_{\alpha\gamma}-\delta_{\alpha'\gamma})J_{\alpha\alpha'}\quad, &
\pi\Gamma_\gamma \,=\,{3\pi^2\over 4}\sum_\alpha\mu_{\gamma\alpha}J_{\gamma\alpha}J_{\alpha\gamma}\quad,
\end{align}
where $J_{\alpha\alpha'}=J^*_{\alpha'\alpha}$ are the initial exchange couplings in the 
interaction part 
$H_{ex}=\sum_{\alpha\alpha'}\,J_{\alpha\alpha'}\,\underline{s}_{\alpha\alpha'}\,\underline{S}$
of the Hamiltonian, where $\underline{s}_{\alpha\alpha'}={1\over 2}\sum_{\sigma k,\sigma' k'}
a^\dagger_{\alpha\sigma k}\,\underline{\sigma}_{\sigma\sigma'}\,a_{\alpha'\sigma' k'}$.

The full RG equations at finite voltages and temperature are quite involved and
will be discussed in detail in a forthcoming work \cite{pletyukhov_hs_reininghaus_preprint}.
Here we consider the equations either for $\mu_\alpha=0$ or for $T=0$,
which simplifies them considerably. Furthermore, we have taken the special
case of two reservoirs with $\mu_L=-\mu_R={V\over 2}$ and used the special initial
condition $J_{\alpha\alpha'}=2\sqrt{x_\alpha x_{\alpha'}}J$ with $0<x_\alpha<1$ and $x_L+x_R=1$
(called ``$2$-reservoir case'' in the following). In this case the conductance 
in units of $G_0={2e^2\over h}$ follows from $G=G(E=0)$ with 
$G(E)=\pi{\partial\over\partial V}\Gamma_L(E)$. 
In the general case, the variation of the current flowing in reservoir $\gamma$ is given
in Laplace space by $\delta I_\gamma(E)={2e\over h}{i\over E}\pi\delta\Gamma_\gamma(E)$. The
spin dynamics follows from 
$\langle\underline{S}\rangle(E)={i\over E+i\Gamma(E)}\langle\underline{S}\rangle(t=0)$.

For zero temperature $T=0$ but arbitrary nonzero $\mu_\alpha$ a lengthy but straightforward 
calculation gives the following RG equations 
\begin{align}
\label{eq:rg_gamma_V}
{\partial^2\over \partial E^2}\Gamma(E) &\,=\, i\Pi_{12} J_{12}(E)J_{21}(E_{12})\quad,\quad
{\partial\over \partial E}\delta\Gamma(E) \,=\, i\delta\mu_{12}\Pi_{12} J_{12}(E)J_{21}(E_{12}),\\
\label{eq:rg_cond_V}
{\partial\over \partial E}\left(\pi\delta\Gamma_\gamma(E)\right) &\,=\, 
-{3\pi^2\over 4}\Pi_{12} I^\gamma_{12}(E)K_{21}(E)(\delta\mu_{12}+iZ_{12}\delta\Gamma_{12}),\\
\nonumber
{\partial\over \partial E}J_{12}(E) &\,=\, 
-{1\over 2}\Pi_{13} J_{13}(E)J_{32}(E_{13})
-{1\over 2}\Pi_{32} J_{13}(E_{32})J_{32}(E)\\
\label{eq:rg_J_V}
&\hspace{1cm}
-{1\over 4}J_{12}(E_{34})J_{34}(E)J_{43}(E_{1234})(Z_{34}\Pi_{1234}+Z_{1234}\Pi_{34}),\\
\nonumber
{\partial\over \partial E}K_{12}(E) &\,=\, -\Pi_{23} J_{13}(E_{21})K_{32}(E)
-\Pi_{31} K_{13}(E)J_{32}(E_{21})\\
\label{eq:rg_K_V}
&\hspace{1cm}
+{1\over 2}J_{12}(E_{2134})J_{34}(E_{21})K_{43}(E)(Z_{34}\Pi_{2134}+Z_{2134}\Pi_{34}),\\
\nonumber
{\partial\over \partial E}I^\gamma_{12}(E) &\,=\, -\Pi_{13} I^\gamma_{13}(E)J_{32}(E_{13})
-\Pi_{32} J_{13}(E_{32})I^\gamma_{32}(E)\\
\label{eq:rg_I_V}
&\hspace{1cm}
+{1\over 2}J_{12}(E_{34})I^\gamma_{34}(E)J_{43}(E_{1234})(Z_{34}\Pi_{1234}+Z_{1234}\Pi_{34}),
\end{align}
where $Z_{12}=Z(E_{12})$, $Z(E)={1\over 1+i{d\Gamma\over dE}}$, $\delta\Gamma_{12}=\delta\Gamma(E_{12})$, 
$\Pi_{12}=\Pi(E_{12})$ and $\Pi_{1234}=\Pi(E_{1234})$, with $\Pi(E)={1\over E+i\Gamma(E)}$.
These equations are even valid for an arbitrary number of reservoirs
and arbitrary initial matrix $J_{\alpha\alpha'}$. We emphasize again
that in these equations the index $1\equiv \alpha_1$ contains only the reservoir label, and
the new notations $E_{12}=E+\mu_{12}$, $\mu_{12}=\mu_1-\mu_2$, and
$E_{1234}=E+\mu_{12}+\mu_{34}$ have been introduced.

For zero voltages $\mu_\alpha=0$ we consider the ``$2$-reservoir case''
and parameterize the renormalized vertices $J_{12}(E)=2\sqrt{x_1 x_{2}}J(E)$,
$K_{12}(E)=2\sqrt{x_1x_2}K(E)$ and 
$I^\gamma_{12}(E)=(\delta_{\alpha_1\gamma}-\delta_{\alpha_2\gamma})2\sqrt{x_1x_2}J_I(E)$.
Omitting the common $E$-argument of all quantities, we obtain the following RG equations
\begin{align}
\label{eq:rg_gamma_J_T_2}
{\partial^2\Gamma\over \partial E^2} &\,=\, 4iJ^2 (F^{(1)}-4JF^{(2)}), &
{\partial J\over \partial E} &\,=\, -2J^2(F^{(3)}+ZJF^{(1)}-2JF^{(4)}), \\
\label{eq:rg_K_JI_T_2}
{\partial K\over \partial E} &\,=\, -4KJ(F^{(3)}-ZJF^{(1)}+2JF^{(4)}), &
{\partial J_I\over \partial E} &\,=\, -2J_IJF^{(3)}, \\
\label{eq:rg_cond_T_2}
{\partial G\over \partial E} &\,=\, 
-\,6\pi^2 x_L x_R\,J_I K (F^{(1)}-6JF^{(2)}),
\end{align}
where we have defined the integrals
\begin{align}
\label{eq:F1}
F^{(1)}(E) &\,=\, Z(E)\int d\omega \int d\omega' {f'(\omega)f'(\omega')\over \omega+\omega'+\chi(E)} 
\,=\, Z(E) \sum_{n,m=0}^{\infty} \frac{2 (2 \pi i T)^2 }{(i \omega_n + i \omega_m + \chi(E))^3}  
\,=\, \frac{Z(E)}{2 \pi i T} \frac{d^2}{d \rho^2} \left[ \rho \psi (\rho) \right],\\
\nonumber
F^{(2)}(E) &\,=\, Z(E)\int d\omega \int d\omega' 
{{\cal{F}}(E,\omega)f'(\omega)f'(\omega')\over \omega+\omega'+\chi(E)}\\
\label{eq:F2}
&= -\frac{Z^2(E)}{2 \pi i T} \left\{ \left[ 1+ \psi (\rho +\textstyle{\frac12}) \right] 
\frac{d^2 }{d \rho^2}\left[ \rho \psi (\rho) \right] -\frac12 \frac{d^2}{d \rho^2} 
\left[ \rho \psi (\rho) \psi (\rho+1)   \right]\right\}, \\
\label{eq:F3}
F^{(3)}(E) &\,=\, -Z(E)\int d\omega {f'(\omega)\over \omega+\chi(E)} 
\,=\,  Z(E) \sum_{n=0}^{\infty} \frac{2 \pi i T}{(i \omega_n +\chi(E))^2}
\,=\, \frac{Z(E)}{2 \pi i T} \frac{d}{d \rho} \psi (\rho +\textstyle{\frac12}),
\\
\label{eq:F4}
F^{(4)}(E) &\,=\, Z(E)\int d\omega {{\cal{F}}(E,\omega)f'(\omega)\over \omega+\chi(E)} 
\,=\, \frac{Z^2(E)}{2 \pi i T} \left\{ \frac12 \frac{d}{d \rho} 
\left[ \psi^2 (\rho +\textstyle{\frac12}) \right] + 
\sum_{n=0}^{\infty} \frac{d }{d \rho} \left[ \frac{\psi (n+1+\rho)}{n+\textstyle{\frac12} +\rho}\right] \right\},
\end{align}
and
\begin{align}
\nonumber
{\cal{F}}(E,\omega) \,&=\, Z(E)\int d\omega'\,f^a(\omega')
\left({1\over\omega+\omega'+\chi(E)}-{1\over\omega'+\chi(E)}\right)\\
\label{eq:H}
&= - Z(E) \,\, 2 \pi i T \sum_{n=0}^{\infty} 
\left( \frac{1}{\omega + i \omega_n + \chi (E)} - \frac{1}{i \omega_n + \chi (E)} \right),
\end{align}
with $f^a(\omega)=f(\omega)-{1\over 2}=-T\sum_n{1\over \omega-i\omega_n}$ 
and the Matsubara frequencies $\omega_n =2 \pi T (n +\frac12)$. As seen above, all these integrals can be expressed 
through the digamma function $\psi$ of the argument $\rho = \frac{\chi(E)}{2 \pi i T}$.

In the following sections we will analyze these RG equations analytically in certain
regimes. Furthermore, we will explain how to set up the initial conditions in order to 
enable an achievement of the scaling limit.

\subsection{T=V=0}
\label{sec:T=V=0}
At $T=V=0$ we use the RG equations (\ref{eq:rg_gamma_J_T_2})-(\ref{eq:rg_cond_T_2}) together
with $F^{(1)}=F^{(3)}={1\over E+i\Gamma(E)}$ and $F^{(2)}=F^{(4)}=0$. With $E=i\Lambda$ and
$\lambda=\Lambda+\Gamma(i\Lambda)$ we obtain
\begin{align}
\label{eq:rg_TV=0}
d^2_\Lambda\Gamma &=-{4J^2\over\lambda}, &
d_\Lambda J &= -{2J^2(1+ZJ)\over\lambda}, &
d_\Lambda K &= -{4KJ(1-ZJ)\over\lambda}, &
d_\Lambda J_I &= -{2J_IJ\over\lambda}, &
d_\Lambda G &= -4x_L x_R{3\pi^2\over 2}{J_I K\over \lambda},
\end{align}
where $Z={1\over 1+d_\Lambda\Gamma}$. With $\tilde{J}=ZJ$, $\tilde{J}_I=ZJ_I$ and 
$l=\ln{{\lambda_0\over\lambda}}$ the equations can be written as
\begin{align}
\label{eq:rg_tilde_TV=0}
{dZ\over dl} &= -4Z\tilde{J}^2, &
{d\tilde{J}\over dl} &= 2\tilde{J}^2(1-\tilde{J}), &
{dK\over dl} &= 4K\tilde{J}(1-\tilde{J}), &
{d\tilde{J}_I\over dl} &= 2\tilde{J}_I\tilde{J}(1-2\tilde{J}), &
{dG\over dl} &= 4x_L x_R{3\pi^2\over 2}\tilde{J}_I K.
\end{align}
These equations have the following invariants which are fixed as follows
\begin{align}
\label{eq:rg_tilde_invariants_TV=0}
T_K &= \lambda e^{-{1\over 2\tilde{J}}}{\tilde{J}^{1/2}\over(1-\tilde{J})^{1/2}}\quad, &
1 &= {Z\over (1-\tilde{J})^2}\quad, &
2 &= {K \over \tilde{J}^2}\quad, &
1 &= {\tilde{J}_I \over \tilde{J}(1-\tilde{J})}\quad.
\end{align}
Hereby $T_K$ sets the energy scale and defines the Kondo temperature.
The other invariants are fixed by comparison with the initial conditions (\ref{eq:initial})
in the scaling limit $\Lambda\rightarrow\infty$ and $\tilde{J}\rightarrow 0$ such that
$T_K$ is kept constant. The initial conditions (\ref{eq:initial}) are given by 
$Z=1$, $K=2J^2$, and $J_I=J$, 
which, in the scaling limit, lead to the invariants (\ref{eq:rg_tilde_invariants_TV=0}). 
Using the results for $\tilde{J}_I$ and $K$, the RG equation for the conductance
can be integrated to
\begin{equation}
\label{eq:rg_cond_TV=0}
G\,=\,4 x_L x_R {3\pi^2\over 4}\tilde{J}^2\quad,
\end{equation}
where the integration constant has been fixed again by comparison to the
initial condition (\ref{eq:initial}).
Having fixed the invariants all information about the microscopic details
of the model parameters is swept into the single energy scale $T_K$.
The only remaining differential equation to be solved
numerically is $d_\Lambda\lambda={1\over Z}={1\over(1-\tilde{J})^2}$, where 
$\tilde{J}$ is a function of $\lambda$ via the expression for $T_K$.
To solve it we need the initial value $\lambda(0)=\Gamma(0)\equiv\bar{\Gamma}$ or, equivalently,
the initial value $\tilde{J}(0)\equiv\bar{J}$. 
This value is determined from the  condition that the normalized conductance is unitary 
$G(0)/(4x_L x_R)=1$ at $T=V=0$ for the $1$-channel Kondo model. This yields the values 
$\bar{J}={2\over \pi\sqrt{3}}\approx 0.37$ and $\bar{\Gamma}\approx 5.11 \,T_K$.  
>From the numerical solution we obtain all quantities at some high energy scale $\Lambda_0\gg T_K$.
At finite $T$ or $\mu_{\alpha}$, we take these values as initial condition at
$\Lambda_0\gg T,|\mu_\alpha|,T_K$ and solve the RG equations starting from $\Lambda_0$.

In second order truncation we neglect the terms $\sim J^3, KJ^2$ on the r.h.s. of (\ref{eq:rg_TV=0})
and obtain $T_K=\lambda J e^{-{1\over 2J}}$, $Z={1\over 1+2J}$, $K=2J^2$, $J_I=J$ and
$G=4x_L x_R{3\pi^2\over 4}J^2$. Fixing unitary conductance yields 
$\bar{J}=J(0)={2\over \pi\sqrt{3}}\approx 0.37$ and 
$\bar{\Gamma}=\Gamma(0)={1\over \bar{J}}e^{1\over 2\bar{J}}\approx 10.57 \,T_K$ for the
initial values for $\Gamma$ and $J$ at $\Lambda=0$.

\emph{Weak coupling.}---To obtain the spin relaxation rate $\Gamma(i\Lambda)$ in the
weak coupling regime $\Lambda\gg T_K$, we expand $\tilde{J}$ and $\Gamma$
around the $2$-loop poor man scaling solution $J_p(\Lambda)$, defined by
$\ln{{\Lambda\over T_K}}={1\over 2J_p}+{1\over 2}\ln{{1-J_p\over J_p}}$.
Using $\Gamma\sim  \Lambda O(J_p)$ and $\tilde{J}=J_p+O(J_p^2)$ we find 
$\lambda=\Lambda(1+O(J_p))$ and ${1\over 2\tilde{J}}=
\ln{{\lambda\over T_K}}-{1\over 2}\ln{{1-\tilde{J}\over \tilde{J}}}=
{1\over 2J_p}+O(J_p)$. This can only be fulfilled if $\tilde{J}=J_p+O(J_p^3)$,
from which we obtain $d_\Lambda\lambda={1\over Z}={1\over(1-\tilde{J})^2}=1+2J_p+3J_p^2+O(J_p^3)$. 
Using the ansatz
$\lambda=\Lambda(1+aJ_p+bJ_p^2\ln{J_p}+cJ_p^2+O(J_p^3))$ and 
$d_\Lambda J_p = -2J_p^2(1-J_p)/\Lambda$, we find 
$d_\Lambda\lambda=1+aJ_p+bJ_p^2\ln{J_p}+(c-2a)J_p^2+O(J_p^3)$. Comparing the
two expressions yields $a=2$, $b=0$ and $c=2a+3=7$, i.e. with $E=i\Lambda$
\begin{align}
\label{eq:gamma_wc}
\tilde{J}(E) &\,=\, J_p(-iE)\,+\,O(J_p^3), &
\Gamma(E) &\,=\, -2iE\left\{J_p(-iE)\,+\,{7\over 2}J_p^2(-iE)\,+\,O(J_p^3)\right\}.
\end{align}

\emph{Time evolution.}---The spin dynamics is calculated from the inverse Laplace transform
$\langle\underline{S}\rangle(t)={1\over 2\pi}\int dE {e^{-iEt}\over \lambda(E)}
\langle\underline{S}\rangle(0)$, with $\lambda(E)=-iE+\Gamma(E)$, where the integration contour lies slightly above
the real axis. The resolvent has a pole at $\lambda(E)=0$, which corresponds
to the fixed point $\tilde{J}(E)=J^*=1$ since 
$\lambda=T_Ke^{-{1\over 2\tilde{J}}}(1-\tilde{J})^{1/2}/\tilde{J}^{1/2}$. 
The fixed point is reached for 
$-iE=\Lambda^*=-\Gamma^*<0$ since at $\Lambda=0$ we have chosen $\tilde{J}(0)={2\over\pi\sqrt{3}}<1$.
Close to the fixed point $\Lambda\sim\Lambda^*=-\Gamma^*$ we get 
${dZ\over d\lambda}=4Z\tilde{J}^2/\lambda\approx\alpha Z/\lambda$
with $\alpha=4{J^*}^2$. This leads to $Z\sim \lambda^\alpha$. Using 
${d\lambda\over d\Lambda}={1\over Z}\sim \lambda^{-\alpha}$ we get
$\lambda\sim (\Lambda-\Lambda^*)^{1 \over 1+\alpha}$.
This means that the resolvent ${1\over\lambda(E)}$ has a branching pole at 
$E=-i\Gamma^*$ followed by a branch cut on the line $E=-i(\Gamma^*+x)$ with $0<x<\infty$. The 
contour integral can be closed in the lower half around the branch cut and,
in the long-time limit $t\gg{1\over \Gamma^*}$, the spin dynamics follows from
\begin{equation}
\langle\underline{S}\rangle(t)
={1\over\pi} e^{-\Gamma^*t}\int_0^\infty dx \,e^{-xt}\text{Im}{1\over\lambda(i(\Lambda^*-x-i\eta))}
\langle\underline{S}\rangle(0)
\sim e^{-\Gamma^*t}\int_0^\infty dx \,e^{-xt}{1\over x^{1\over 1+\alpha}}
\sim {1\over t^g}\,e^{-\Gamma^*t},
\end{equation}
with the exponent
\begin{equation}
\label{eq:exponent}
g\,=\,{\alpha\over 1+\alpha}={4{J^*}^2\over 1+4{J^*}^2}.
\end{equation}
As a result the exponent depends on the fixed point value $J^*$, which in turn depends on the
truncation order. In third order truncation we get $J^*=1$ leading to $g={4\over 5}$. 
In second order truncation we neglect the term of $O(J^3)$ in the RG equation (\ref{eq:rg_TV=0}) for $J$. This
yields ${d\tilde{J}\over dl}=2\tilde{J}^2(1-2\tilde{J})$ and $Z=1-2\tilde{J}$.
The fixed point in second order truncation is at $\tilde{J}=J^*={1\over 2}$ which yields $g={1\over 2}$.
A fixed point value $J^*=\infty$ would lead to $g=1$. This can be achieved 
by changing the $\beta$-function by hand to ${d\tilde{J}\over dl}={2\tilde{J}^2\over 1+\tilde{J}}$,
which is still consistent with the $\beta$-function in third order truncation.
This leads to the invariant $T_K=\lambda\tilde{J}^{1/2}e^{-{1\over 2\tilde{J}}}$
and to $Z=e^{-2\tilde{J}-\tilde{J}^2}$. Close to the fixed point $\lambda=0$, the
coupling $\tilde{J}\sim{1\over\lambda^2}$ tends to infinity and fulfills the 
differential equation 
$d_\Lambda\tilde{J}=-{1\over Z\lambda}{2\tilde{J}^2\over 1+\tilde{J}}\sim 
-\tilde{J}^{3/2}e^{2\tilde{J}+\tilde{J}^2}$. In the leading order we get 
$\tilde{J}~\sim (-\ln{\Lambda-\Lambda^*\over T_K})^{1/2}$ and 
$\lambda\sim \tilde{J}^{-1/2}\sim (-\ln{\Lambda-\Lambda^*\over T_K})^{-1/4}$,
which results in a pre-exponential function $\sim {1\over t(\ln{T_K t})^{3/4}}$. Thus,
besides the power-law ${1\over t}$ the pre-exponential function contains an
additional logarithmic term in this case. In conclusion, the pre-exponential 
modulation depends crucially on the form of the $\beta$-function
and can not be determined within a perturbative truncation scheme. Instead,
one needs a non-perturbative analysis of the $\beta$-function close to the
fixed point which goes beyond the scope of this work.

The problem of obtaining the correct pre-exponential modulation for the $1$-channel Kondo
model is analogous to the problem of obtaining the correct scaling dimension close to the fixed point 
for multi-channel Kondo models. For channel number $N>1$ the RG equations for $Z$ and $\tilde{J}$ change to
${dZ\over dl}=-4NZ\tilde{J}^2$ and ${d\tilde{J}\over dl}=2\tilde{J}^2(1-N\tilde{J})$,
giving $T_K=\lambda e^{-{1\over 2\tilde{J}}}({\tilde{J}\over 1-N\tilde{J}})^{N/2}$
and $Z=(1-N\tilde{J})^2$. 
The fixed point $\lambda=0$ corresponds to $\tilde{J}={1\over N}$ and the system should reveal 
non-Fermi liquid behavior at low energies \cite{nozieres,cft}. This means that we have 
to choose $\tilde{J}(E=0)={1\over N}$ identical to the fixed point value such
that the pole $\lambda=0$ lies at the origin $E=\Lambda^*=0$ of the complex plane
with $\Gamma(0)=\Gamma^*=0$. In contrast to the exponential decay for the $1$-channel case, 
this leads to a pure power-law decay in the long-time limit, which is another signature of
non-Fermi liquid behavior. Expanding around the fixed point we get 
${dZ\over d\lambda}\approx\alpha Z/\lambda$ with $\alpha=4N{J^*}^2$. According to
(\ref{eq:exponent}) this gives
the power-law exponent $g={\alpha\over 1+\alpha}={4N{J^*}^2\over 1+4N{J^*}^2}$.
Inserting $J^*={1\over N}$ we obtain $g={4\over N+4}$.
However, this result is only consistent for $N\gg 1$, where the $\beta$-function
can be systematically truncated and the system stays in the weak-coupling limit \cite{mitra_rosch_PRL11}.
For small $N$ the result is not reliable like in the $1$-channel case discussed above. 

\subsection{T=0 and finite voltage}
\label{sec:finite_V}
At zero temperature we start from the RG equations (\ref{eq:rg_gamma_V}-\ref{eq:rg_I_V})
and discuss analytically the ``$2$-reservoir case'' in the weak coupling regime of large 
voltage $V\gg T_K$ and the strong coupling regime of small voltage $V\ll T_K$.

\emph{Weak coupling.}---To obtain the weak coupling expansion for $V\gg T_K$ we follow 
Ref.~\onlinecite{hs_reininghaus_PRB09} and discuss the solution of the RG equations
separately for $\Lambda>V$ and $\Lambda<V$, where $\Lambda=-iE$. For $\Lambda>V$, we 
expand the solution of the RG equations
around a reference solution, where all propagators are replaced by ${Z\over \Lambda}$.
No voltage dependence is generated in this case and the RG equations
obtain the same form as in (\ref{eq:rg_tilde_TV=0}) but with $l\rightarrow\ln{\Lambda_0\over\Lambda}$.
As a consequence, the reference solution is given by (\ref{eq:rg_tilde_invariants_TV=0}) but
with $\tilde{J}\rightarrow J_p$, where $J_p(\Lambda)$ is the $2$-loop poor man scaling solution,
already introduced in section~\ref{sec:T=V=0}. To calculate the first correction to the reference
solution for $\Lambda>V$, we insert the exact identity
\begin{align}
\nonumber
{1\over \Lambda_{12}+\Gamma_{12}}H_\Lambda \,&=\, {Z\over \Lambda}H_\Lambda \,+\,
{\partial\over \partial \Lambda}\left\{Z_{12}(\ln{\Lambda_{12}+\Gamma_{12}\over \Lambda})H_\Lambda\right\}\\
\label{eq:wc}
\,&-\,(\ln{\Lambda_{12}+\Gamma_{12} \over \Lambda}){\partial\over \partial \Lambda}\left\{Z_{12}H_\Lambda\right\}
\,+\,{Z_{12}-Z \over \Lambda}H_\Lambda
\end{align}
for all propagators occurring in the RG equations (\ref{eq:rg_gamma_V}-\ref{eq:rg_I_V}), where 
$\Lambda_{12}=\Lambda-i\mu_{12}$ and $H_\Lambda$
stands symbolically for the rest of some term on the r.h.s. of the RG equation involving products
of various couplings. We get $H_\Lambda\sim J_p^2$ ($H_\Lambda\sim J_p^3$) for 
$d_\Lambda J$,$d_\Lambda I^\gamma$ and $d_\Lambda Z$ ($d_\Lambda K$ and $d_\Lambda G$, with 
$G=\pi{\partial\over\partial V}\Gamma_L$). The first term of (\ref{eq:wc})
generates the reference solution, which, according to (\ref{eq:rg_tilde_invariants_TV=0}) and 
(\ref{eq:rg_cond_TV=0}), gives 
$O(J_p)$ ($O(J_p^2)$) for $J$, $I^\gamma$ and $Z-1$ ($K$ and $G$). We are aiming at calculating
all quantities one order beyond the reference solution. To estimate the various terms of 
(\ref{eq:wc}) we use 
$\ln{\Lambda_{12}+\Gamma_{12} \over \Lambda}\sim O({V\over\Lambda},{\Gamma\over\Lambda})
\sim O({V\over\Lambda},J_p)$ for $\Lambda\gg V,\Gamma$, where we used $\Gamma\sim O(\Lambda J_p)$ 
according to (\ref{eq:gamma_wc}). As a consequence, the third term of (\ref{eq:wc}) is 
$\sim O({V\over \Lambda^2}J_p^3,{1\over\Lambda}J_p^4)$ 
($\sim O({V\over \Lambda^2}J_p^4,{1\over\Lambda}J_p^5)$)
for $d_\Lambda J$,$d_\Lambda I^\gamma$ and $d_\Lambda Z$ ($d_\Lambda K$ and $d_\Lambda G$). 
Using ${\partial Z\over \partial \Lambda}\sim {1\over \Lambda}O(J_p^2)$, the fourth term 
of (\ref{eq:wc}) is $\sim O({V\over \Lambda^2}J_p^4)$ ($\sim O({V\over \Lambda^2}J_p^5)$).
Thus, after integrating over $\Lambda$, the third and fourth terms of (\ref{eq:wc}) produce
corrections $\sim J_p^3$ ($\sim J_p^4$) for $J$, $I^\gamma$ and $Z$ ($K$ and $G$), i.e. two orders higher
than the reference solution. In contrast, the second term of (\ref{eq:wc}) produces the first 
correction $\sim{V\over \Lambda}O(J_p^2)$ ($\sim{V\over \Lambda}O(J_p^3)$) to the reference solution. 
We can neglect the influence of this correction on the first term of (\ref{eq:wc}), since this
gives contributions of the same order as the third term of (\ref{eq:wc}). Thus for our purpose it is 
sufficient to take into account the first two terms of (\ref{eq:wc}) and 
calculate $H_\Lambda$ by using the reference solution. For the RG equation (\ref{eq:rg_cond_V}) for
the conductance this means that the term involving $\delta\Gamma_{12}\sim \delta V J_p$ can be neglected on the r.h.s.
and the first term on the r.h.s. leads to the first correction to the reference solution, i.e. the weak coupling 
expansion in the regime $\Lambda>V$ reads
\begin{equation}
\label{eq:wc_cond_Lambda>V}  
G(E=i\Lambda,V)\,=\,4 x_L x_R{3\pi^2\over 4}J_p^2 (\Lambda)
\left\{1\,-\,2J_p(\Lambda)\sum_{\alpha_1\ne\alpha_2}\ln{\Lambda_{12}+i\Gamma_{12}\over \Lambda}\,+\,O(J_p^2)\right\}.
\end{equation}
Following Ref.~\onlinecite{hs_reininghaus_PRB09} we use this result at $\Lambda=V$ as 
initial condition to solve the RG equation for the
conductance in the following regime $\Lambda<V$ by expanding in the reference solution $J_p(V)$
evaluated at $\Lambda=V$. Replacing all vertices on the r.h.s. of the RG equation for the
conductance by the reference solution at $\Lambda=V$, we get with $H_V\sim J_I K \sim O(J_p^3)$ and 
${\partial Z\over\partial\Lambda}\sim {1\over \Lambda+\Gamma}O(J_p^2)$ 
instead of (\ref{eq:wc}) the expression
${1\over \Lambda_{12}+\Gamma_{12}}H_V =
{\partial\over \partial \Lambda}\left\{Z_{12}(\ln{\Lambda_{12}+\Gamma_{12}\over V})H_V\right\}
+{1\over\Lambda+\Gamma}O(J_p^5)$. Applying $\int_V^0 d\Lambda$ to this expression and
adding the initial condition (\ref{eq:wc_cond_Lambda>V}) evaluated at $\Lambda=V$, the 
second term on the r.h.s. of (\ref{eq:wc_cond_Lambda>V}) obviously cancels against the
contribution from the lower bound $\Lambda=V$ of the integration. The remaining first term of 
(\ref{eq:wc_cond_Lambda>V}) evaluated at $\Lambda=V$ together with the contribution from 
the upper bound $\Lambda=0$ of the integration
gives for the conductance at $E=0$ the result $G=4 x_L x_R{3\pi^2\over 4}J_p^2 (V)
\left\{1-2J_p(V)\sum_{\alpha_1\ne\alpha_2}\ln{-i\mu_{12}+i\Gamma(\mu_{12})\over V}+O(J_p^2)\right\}$, where we used
$Z_{12}|_{\Lambda=0}=1+O(J_p(V))$. With $\mu_{12}=\alpha_{12}{V\over 2}$ and $\alpha_{12}=\alpha_1-\alpha_2$, we 
find 
\begin{equation}
\sum_{\alpha_1\ne\alpha_2}\ln{-i\mu_{12}+i\Gamma(\mu_{12})\over V}=
\sum_{\alpha_1\ne\alpha_2}\ln(-i{1\over 2}\alpha_{12})+O({\Gamma\over V})
=\ln(-i)+\ln(i)+O({\Gamma\over V})=O({\Gamma\over V})\sim J_p^2,
\end{equation}
where we have used $\Gamma(E=0)\sim O(VJ_p^2)$ which can be derived from an
analogous weak coupling expansion. Thus, we can neglect this term and finally get the weak coupling result
\begin{equation}
\label{eq:wc_cond}  
G(V)\,=\,4 x_L x_R{3\pi^2\over 4}J_p^2 (V) \,+\,O(J_p^4).
\end{equation}

\emph{Strong coupling.}---In the strong coupling regime $V \ll T_K$ the 1-channel Kondo model 
behaves like a local Fermi liquid \cite{nozieres}, which, in particular, implies a quadratic 
dependence of the differential conductance on the applied bias voltage 
\cite{oguri_JPSJ05,sela_malecki_PRB09} $G=4x_L x_R \{1-c_V(V/\bar{\Gamma})^2+O((V/\bar{\Gamma})^3) \}$. 
As a reference scale we use here $\bar{\Gamma}=\Gamma(E=0)$. In the following we recover it by means of the 
RTRG as well as provide a quantitative value for the Fermi-liquid coefficient $c_V$ 
up to the next-to-leading order in the renormalized coupling constant 
$\bar{J}=\tilde{J}(E=0)|_{T=V=0}$. This expansion can be performed by differentiating
the RG equation (\ref{eq:rg_cond_V}) for the normalized conductance twice by the voltage
and integrate it over $E$. A more direct path is to consider the original diagrammatic
series and calculate directly the third order variation 
$\delta^3\Sigma_\gamma(E)={1\over 3!}{\partial^3\over {\partial V}^3}\Sigma_\gamma(E){\delta V}^3$
of the current kernel by using the diagrammatic method explained in Section~\ref{sec:E_flow}.
Of particular advantage is the fact that the variation $\delta\Gamma(E)$ of the Liouvillian
vanishes at $V=T=0$ for any $E$. This follows directly from the RG equation (\ref{eq:rg_gamma_V})
after inserting the algebra of the vertices at $V=T=0$ derived in Section~\ref{sec:T=V=0}, given
by 
\begin{align}
\nonumber
J_{\alpha_1\alpha_2}&=2\sqrt{x_{\alpha_1} x_{\alpha_2}}\,J, & 
K_{\alpha_1\alpha_2}&=2\sqrt{x_{\alpha_1}x_{\alpha_2}}\,K, & 
I^\gamma_{\alpha_1\alpha_2}&=(\delta_{\alpha_1\gamma}-\delta_{\alpha_2\gamma})
2\sqrt{x_{\alpha_1}x_{\alpha_2}}\,J_I ,\\ 
\label{eq:vertices_T=V=0}
Z&=(1-\tilde{J})^2, & K&=2\tilde{J}^2, & \tilde{J}_I&=\tilde{J}(1-\tilde{J}),
\end{align}
where $\tilde{J}=ZJ$ and $\tilde{J}_I=ZJ_I$. As a consequence we can express the variation
of the propagator at $V=0$ by derivatives with respect to the Laplace variable
$\delta^n\Pi_{12}={1\over n!}(\delta\bar{\mu}_{12})^n\partial_E^n\Pi_{12}$. With this property we
get for the third order variation of the current kernel at zero voltage 
\begin{align}
\nonumber
\delta^3\Sigma_\gamma(E)|_{V=0}\quad &= \quad
{1\over 2}{1\over 3!}(\delta\bar{\mu}_{12})^3
\begin{picture}(10,10)
\put(5,-13){\includegraphics[height=1cm]{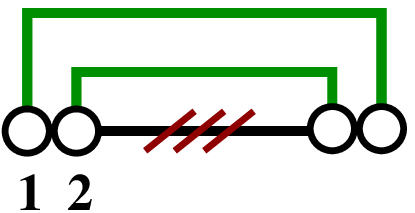}}
\end{picture}
\\ \nonumber \\
\label{eq:Sigma_variation}
&\hspace{-1cm}+\quad
{1\over 2}(\delta\bar{\mu}_{12})^2\delta\bar{\mu}_{13}
\begin{picture}(10,10)
\put(5,-10){\includegraphics[height=0.8cm]{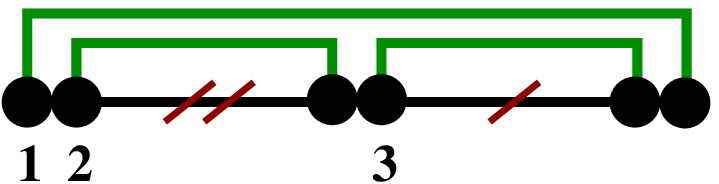}}
\end{picture}
\hspace{3.5cm}+\quad
{1\over 2}\delta\bar{\mu}_{12}(\delta\bar{\mu}_{13})^2
\begin{picture}(10,10)
\put(5,-10){\includegraphics[height=0.8cm]{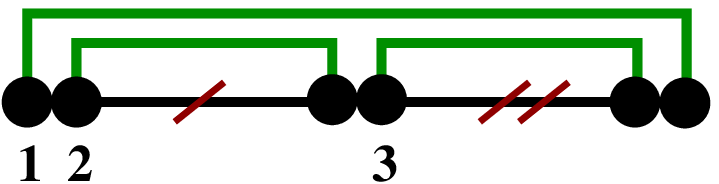}}
\end{picture}
\hspace{3.5cm}+\quad O(\tilde{J}^5),
\end{align}
where all diagrams have to be evaluated at $V=0$ on the r.h.s. The corrections
are of $O(\tilde{J}^5)$ since for the current kernel one of the vertices has to
be $K\sim\tilde{J}^2$. At $V=0$ the last two diagrams are identical and, 
by using partial integration, can be written as
\begin{align}
\label{eq:identity}
\begin{picture}(10,10)
\put(5,-10){\includegraphics[height=0.8cm]{L_cV_2.eps}}
\end{picture}
\hspace{3cm}\quad=\quad
\begin{picture}(10,10)
\put(5,-10){\includegraphics[height=0.8cm]{L_cV_3.eps}}
\end{picture}
\hspace{3cm}\quad 
=\quad -{1\over 2}
\begin{picture}(10,10)
\put(5,-10){\includegraphics[height=1cm]{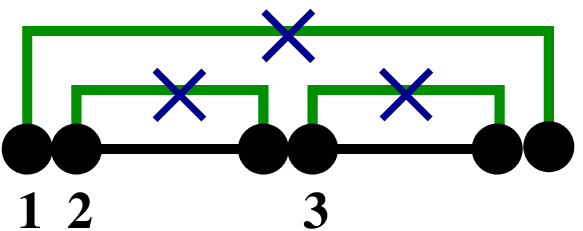}}
\end{picture}
\hspace{3cm}
\end{align}
Analogously to (\ref{eq:L_partial}) we can perform two partial integrations
with the first term on the r.h.s. of (\ref{eq:Sigma_variation}) and, by
again using (\ref{eq:identity}), we arrive at
\begin{align}
\nonumber
{\partial^3\over {\partial V}^3}\Sigma_\gamma(E)|_{V=0}\,\,{\delta V}^3\quad &= \quad
{1\over 2}(\delta\bar{\mu}_{12})^3
\begin{picture}(10,10)
\put(5,-10){\includegraphics[height=1cm]{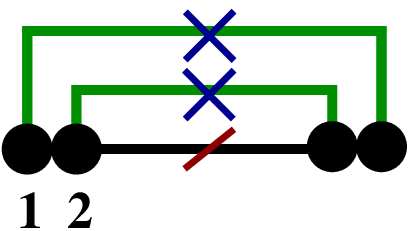}}
\end{picture}
\\ 
\label{eq:Sigma_variation_final}
&\hspace{-1cm}+\quad
{1\over 2}\left\{(\delta\bar{\mu}_{12})^3 + (\delta\bar{\mu}_{13})^3 
-3(\delta\bar{\mu}_{12})^2\delta\bar{\mu}_{13}-3\delta\bar{\mu}_{12}(\delta\bar{\mu}_{13})^2\right\}
\begin{picture}(10,10)
\put(5,-13){\includegraphics[height=1cm]{L_cV_5.eps}}
\end{picture}
\hspace{2.8cm}+\quad O(\tilde{J}^5).
\end{align}
Due to the derivative of the Fermi functions in all contractions, all frequencies
are set to zero and the diagrams can be straightforwardly evaluated by performing
the sums over the indices $\eta$ and $\sigma$ by using the algebra of the vertices together with
their value (\ref{eq:vertices_T=V=0}) at $V=0$. With $\mu_\alpha={1\over 2}\alpha V$,
$\alpha_{12}=\alpha_1-\alpha_2$ and 
$\delta_{\alpha_1\gamma}-\delta_{\alpha_2\gamma}={1\over 2}\gamma\alpha_{12}$, we obtain after
a straightforward calculation
\begin{align}
\nonumber
{\partial^3\over {\partial V}^3}\Sigma_\gamma(E)|_{V=0}\,\,{\delta V}^3\quad &= \quad
i\gamma L^b \,4x_L x_R{\pi\over 64}\,{\tilde{J}_I K(1+3\tilde{J})\over Z^2(E+i\Gamma)^2} \\
\label{eq:Sigma_variation_eval}
&\hspace{-2cm}\times\,\left\{\sum_{\alpha_1\alpha_2}(\alpha_{12})^4 +
\sum_{\alpha_1\alpha_2\alpha_3}\alpha_{12}\left((\alpha_{12})^3 +(\alpha_{13})^3 
-3(\alpha_{12})^2\alpha_{13} -3\alpha_{12}(\alpha_{13})^2\right)\right\}{\delta V}^3
\quad + \quad O(\tilde{J}^5).
\end{align}
Performing the sums over $\alpha_i$ and using $\Sigma_\gamma(E)=i\Gamma_\gamma(E)L^b$
along with $G=\pi{\partial\over\partial V}\Gamma_L$ and 
(\ref{eq:vertices_T=V=0}), we obtain at $E=0$ the final result
\begin{equation}
\label{eq:cV}
c_V\,=\,-{1\over 2}\,{\bar{\Gamma}^2\over 4x_L x_R}{\partial^3\over{\partial V}^3}\pi\Gamma_\gamma(E=0)\,=\,
{3\pi^2\over 2}{\bar{J}^3(1+3\bar{J})\over (1-\bar{J})^3}\,+\,O(\bar{J}^5).
\end{equation}

Finally, we note that the second variation of the current kernel at $V=0$ 
yields zero since an analogous derivation as above gives
\begin{align}
\nonumber
\delta^2\Sigma_\gamma(E)|_{V=0}\quad &= \quad
{1\over 2}{1\over 2}(\delta\bar{\mu}_{12})^2
\begin{picture}(10,10)
\put(5,-13){\includegraphics[height=1cm]{L_Eder_1_ind.eps}}
\end{picture}
\hspace{2.2cm}+\quad
\delta\bar{\mu}_{12}\delta\bar{\mu}_{13}
\begin{picture}(10,10)
\put(5,-13){\includegraphics[height=1cm]{L_Eder_2_ind.eps}}
\end{picture}
\hspace{4.2cm}+\quad O(\tilde{J}^5),
\\ \nonumber \\
\label{eq:Sigma_2nd_variation}
&=\quad
{1\over 4}(\delta\bar{\mu}_{12})^2
\begin{picture}(10,10)
\put(5,-13){\includegraphics[height=1cm]{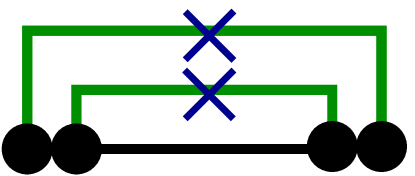}}
\end{picture}
\hspace{2.5cm}-\quad
{1\over 2}(\delta\bar{\mu}_{12}-\delta\bar{\mu}_{13})^2
\begin{picture}(10,10)
\put(5,-13){\includegraphics[height=1cm]{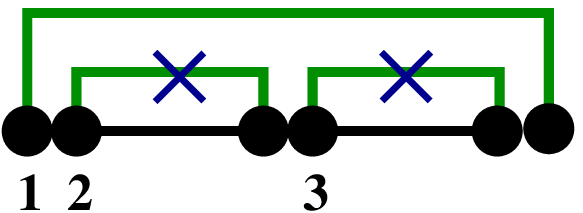}}
\end{picture}
\hspace{3cm}+\quad O(\tilde{J}^5).
\end{align}
Inserting the algebra of the vertices along with their values (\ref{eq:vertices_T=V=0}) 
at $V=0$, and using $\sum_{\alpha_1\alpha_2}(\alpha_{12})^3=0$ and
$\sum_{\alpha_1\alpha_2\alpha_3}(\alpha_{12}-\alpha_{13})^2\alpha_{12}=0$, we find
a vanishing value. This shows that the corrections of the conductance to the unitary
value start with a term $\sim V^2$ as required by the Fermi liquid theory.

\subsection{V=0 and finite temperature}
At zero voltage we consider the ''$2$-reservoir case'' and start from the RG equations 
(\ref{eq:rg_gamma_J_T_2}-\ref{eq:rg_cond_T_2}) to discuss analytically the weak and strong coupling regimes
$T\gg T_K$ and $T\ll T_K$.

\emph{Weak coupling.}---To obtain the weak coupling expansion for $T\gg T_K$ we proceed
analogously to the derivation presented in Section~\ref{sec:finite_V}. We set $E=i\Lambda$ and start with the regime
$\Lambda>T$, where we expand in the reference solution $J_p(\Lambda)$. As already 
discussed in Sections~\ref{sec:finite_V} and \ref{sec:T=V=0}, the reference solution 
(\ref{eq:rg_tilde_invariants_TV=0}) with $\tilde{J}\rightarrow J_p$ is obtained if one
replaces the integrals $F^{(i)}(i\Lambda)$ defined in (\ref{eq:F1}-\ref{eq:F4}) by 
$iF^{(1,3)}\rightarrow {Z\over \Lambda}$ and $F^{2,4}\rightarrow 0$ in
the RG equations (\ref{eq:rg_gamma_J_T_2}-\ref{eq:rg_cond_T_2}). Therefore, we construct
an analog of (\ref{eq:wc}) by writing 
$iF^{(1,3)}={Z\over\Lambda}+Zd_\Lambda \tilde{F}^{(1,3)}+O(J_p^2/\Lambda)$,
such that $\tilde{F}^{(1,3)}\sim O(({T\over\Lambda})^2,J_p)$
for $\Lambda\gg T$. Using $\rho={Z(\Lambda+\Gamma)\over 2\pi T}$
and ${d\rho\over d\Lambda}={1\over 2\pi T}(1+ O(J_p^2))$ together with 
(\ref{eq:F1}) and (\ref{eq:F3}), we find that this can be achieved by defining 
$\tilde{F}^{(1)}={d\over d\rho}\{\rho\Psi(\rho)\}-1-\ln{Z\Lambda\over 2\pi T}$ and
$\tilde{F}^{(3)}=\Psi(\rho+{1\over 2})-\ln{Z\Lambda\over 2\pi T}$. Furthermore, we write 
$iF^{(2,4)}=Zd_\Lambda \tilde{F}^{(2,4)}+O(J_p^2/\Lambda)$  
such that $\tilde{F}^{(2,4)}\sim O({T\over\Lambda})^2$, which follows directly
from the asymptotic expansion $F^{(2,4)}\sim {T^2\over\Lambda^3}$ for $\Lambda\gg T$.
The analog of (\ref{eq:wc}) then reads
\begin{equation}
\label{eq:wc_F}
iF^{(i)}_\Lambda H_\Lambda\,=\,c_i{Z_\Lambda\over\Lambda}H_\Lambda\,
+\,{d\over d\Lambda}\left\{Z_\Lambda\tilde{F}^{(i)}_\Lambda H_\Lambda\right\}\,
-\,\tilde{F}^{(i)}_\Lambda{d\over d\Lambda}\left\{Z_\Lambda H_\Lambda\right\}\,
+\,O({1\over \Lambda} J_p^2 H),
\end{equation}
with $c_1=c_3=1$ and $c_2=c_4=0$. All arguments used in Section~\ref{sec:finite_V}
can then be overtaken in the same way, one just has to replace the logarithmic function
$\ln{\Lambda_{12}+\Gamma_{12}\over \Lambda}$ by the functions $\tilde{F}^{(i)}$ in the
corresponding terms. In particular this means that the terms with the integrals 
$F^{(2,4)}$ do not contribute to the first correction beyond the reference solution
since they appear in third order and fall off sufficiently fast 
$\sim {T^2\over \Lambda^3}$ for $\Lambda\gg T$. Thus, for the conductance, we obtain
by analogy with (\ref{eq:wc_cond_Lambda>V}) for $\Lambda>T$
\begin{equation}
\label{eq:wc_cond_Lambda>T}
G(E=i\Lambda,T)\,=\,4 x_L x_R{3\pi^2\over 4}J_p^2 (\Lambda)
\left\{1\,-\,4J_p(\Lambda)\tilde{F}^{(1)}_\Lambda(T)\,+\,O(J_p^2)\right\}.
\end{equation}
Following Section~\ref{sec:finite_V}, in the next regime $\Lambda<T$, we expand in 
the reference solution $J_p(T)$ evaluated at $T$. Replacing all vertices on the
r.h.s. of (\ref{eq:rg_cond_T_2}) by the reference solution at $\Lambda=T$, we
see that only the first term involving $F^{(1)}$ can correct the result in
$O(J_p^3)$. Furthermore, for the integration $\int_T^0\,d\Lambda$, we use the
identity $iF^{(1)}_\Lambda=Z_{\Lambda=T} d_\Lambda \hat{F}^{(1)}_\Lambda+O({1\over T}J_p^2)$, 
with $\hat{F}^{(1)}_\Lambda={d\over d\rho}\{\rho\Psi(\rho)\}-1-\ln{Z_{\Lambda=T}\over 2\pi}$,
which can be proven by using $|Z_\Lambda-Z_{\Lambda=T}|\sim O(J_p^2)$,
${d\rho\over d\Lambda}={1\over 2\pi T}(1+ O(J_p^2))$
and $F^{(1)}_\Lambda\sim O({1\over T})$ for $\Lambda< T$. This form is
constructed in such a way that $\hat{F}^{(1)}_\Lambda$ is well-behaved for
$\Lambda\rightarrow 0$ and, at the lower boundary $\Lambda=T$ of the integration, is
exactly identical to $\tilde{F}^{(1)}_{\Lambda=T}$, such that the third order
term of (\ref{eq:wc_cond_Lambda>T}) is canceled. The remaining first term of 
(\ref{eq:wc_cond_Lambda>T}) evaluated at $\Lambda=T$ together with the contribution from 
the upper bound $\Lambda=0$ of the integration gives for the conductance at $E=0$ the
result $G=4 x_L x_R{3\pi^2\over 4}J_p^2 (T)
\left\{1-4J_p(T)\hat{F}^{(1)}_{\Lambda=0}+O(J_p^2)\right\}$. 
Using $\Gamma_{\Lambda=0}\sim O(T J_p^2)$ (which can be derived from
an analogous weak coupling expansion) we get 
$\rho_{\Lambda=0}={Z_{\Lambda=0}\Gamma_{\Lambda=0}\over 2\pi T}\sim O(J_p^2)$,
i.e. $\Lambda=0$ is equivalent to $\rho=0$ up to higher-order terms. Using
${d\over d\rho}\{\rho\Psi(\rho)\}_{\rho=0}=-\gamma$,
where $\gamma=0.577\dots$ is Euler's constant, and $Z_{\Lambda=T}=1+O(J_p)$, we 
get $\hat{F}^{(1)}_{\Lambda=0}=-\gamma-1+\ln{(2\pi)}+O(J_p)$, leading to the final 
weak coupling result
\begin{equation}
\label{eq:wc_cond_T}  
G(T)\,=\,4 x_L x_R{3\pi^2\over 4}J_p^2(T)
\left\{1\,-\,J_p(T)\,4\left(\ln{(2\pi)}-1-\gamma\right) \,+\,O(J_p^2)\right\}.
\end{equation}

\emph{Strong coupling.}---In the strong coupling regime $T \ll
T_K$ we aim at deriving the Fermi liquid coefficient $c_T$ in the low
temperature expansion of the conductance $G=4x_L x_R \{ 1-c_T(T/\bar{\Gamma})^2+O((T/\bar{\Gamma})^3) \}$
or of its first derivative 
${\partial G\over \partial T}=-8x_L x_R {1\over \bar{\Gamma}^2} c_T T+O(T^2/\bar{\Gamma}^3)$.
To determine the latter quantity we consider the $T$-derivative of the RG equations 
(\ref{eq:rg_gamma_J_T_2}-\ref{eq:rg_cond_T_2}) and integrate them formally over $E$
(alternatively, one can also derive a diagrammatic technique for $T$-derivatives similar to the one
described in Section~\ref{sec:E_flow} for $E$- and voltage derivatives 
\cite{pletyukhov_hs_reininghaus_preprint}). Employing the representation
$F^{(i)}(E)={Z(E)\over T}H^{(i)}({\chi(E)\over T})$ we find the useful relations
\begin{align}
\label{eq:F_property}
\frac{\partial F^{(1/3)}}{\partial T} &= 
-\frac{Z}{T} \frac{\partial}{\partial E}\left(\frac{\chi F^{(1/3)}}{Z}-1\right) + O (J^2), &
\frac{\partial F^{(2/4)}}{\partial T} &= 
-\frac{Z}{T} \frac{\partial}{\partial E}\left(\frac{\chi F^{(2/4)}}{Z}\right) + O (J^2),
\end{align}
along with the asymptotic behavior $\chi F^{(1,3)}-Z\,,\,\chi F^{(2,4)}\sim O({T^2\over E^2})$
for $E\gg T$. More precisely, for $E\gg T$, the corrections to (\ref{eq:F_property})
are of $O(J^2/E^2)$, such that
they can not lead to a reduction of the order in $J$ when integrating the RG equations
in the regime from $E=\infty$ to $E=T$. This follows from analyzing the result
of the integration procedure, which shows that, for $E\gg T$, we get
${\partial (J,J_I,Z) \over \partial T}\sim {T\over E^2}J^2$, 
${\partial K \over \partial T}\sim {T\over E^2}J^3$,
and ${\partial \Gamma \over \partial T}\sim {T\over E}J^2$. Furthermore, we made use
of ${\partial \chi \over \partial E}=1+O(J^2)$. As a consequence, one finds
for the $T$-derivative of (\ref{eq:rg_K_JI_T_2}) 
\begin{equation}
\label{eq:JIK_T_flow}
{\partial\over\partial E} {\partial (J_I K) \over \partial T} = 
{6J_IKJZ \over T}{\partial\over\partial E}\left({\chi F^{(3)}\over Z}-1\right)\,+\,O(J^5),
\end{equation}
where the corrections are again of $O(J^5/E^2)$ for $E\gg T$.
Integrating (\ref{eq:JIK_T_flow}) over $E$ we obtain to the leading order in $J$
\begin{equation}
\label{eq:JIK}
\frac{\partial (J_I K)}{\partial T} = \frac{6 J_I KJ}{T} 
\left(\chi F^{(3)} - Z \right) + O (J^5).
\end{equation}
Using this result along with (\ref{eq:F_property}) to calculate the 
$T$-derivative of (\ref{eq:rg_cond_T_2}) we obtain
\begin{eqnarray}
\frac{\partial}{\partial E} \frac{\partial G}{\partial T}
&=&  -\frac{6 \pi^2 x_L x_R}{T} \left[ 6 J_I K J \left( \chi F^{(3)}-Z\right) F^{(1)}  
- J_I K Z {\partial\over \partial E} \left( {\chi F^{(1)}\over Z}-1
- 6 J \frac{\chi F^{(2)}}{Z}\right)\right] + O (J^5) \nonumber \\
&=& \frac{6 \pi^2 x_L x_R}{T} \left\{ \frac{\partial}{\partial E}
\left[ J_I K (\chi F^{(1)} -Z) - 6 J_I K J \chi F^{(2)} \right] + 6 J_I K J Z 
\left[ F^{(1)} - F^{(3)}\right]\right\} + O (J^5)\quad,
\label{eq:rg_cond_T_4}
\end{eqnarray}
where we used ${\partial (J_IK)\over \partial E}=-6J_IKJF^{(3)}+O(J^5)$ and 
${\partial Z\over \partial E}\sim O(J^2)$ in the last step.
Using (\ref{eq:F1}) and (\ref{eq:F3}) as well as ${d\rho\over dE}={1\over 2\pi iT}+O(J^2)$, we
can integrate this result over $E$ and obtain the equation
\begin{equation}
\frac{\partial G}{\partial T} =  \frac{6 \pi^2 x_L x_R}{T}  J_I K 
\left[ \left(\chi F^{(1)} -Z \right)- 6 J  \left( \chi F^{(2)} - \bar{F}^{(2)}\right)\right] + O (J^5),
\label{eq:rg_cond_T_5}
\end{equation}
where $\bar{F}^{(2)}=Z^2\left[{d\over d\rho}(\rho\psi(\rho))-\psi(\rho+{1\over 2})-1\right]$
with the asymptotic behavior $\bar{F}^{(2)}\sim O({T^2\over E^2})$ for $E\gg T$.

Setting $E=0$ in (\ref{eq:rg_cond_T_5}) and using the low temperature expansions
$\chi F^{(1)}\rightarrow Z(1-{1\over 6\rho^2})$, $\chi F^{(2)}\rightarrow Z^2{1\over 8\rho^2}$
and $\bar{F}^{(2)}\rightarrow Z^2{1\over 24\rho^2}$ for $\rho={Z(0)\Gamma(0)\over 2\pi T}\gg 1$, we get
\begin{equation}
\frac{\partial G}{\partial T} =  -\frac{4 \pi^4 x_L x_R \bar{J}_I \bar{K}}{(\bar{Z} \bar{\Gamma})^2}  
\left( 1+3 \bar{J}\right)\,T + O(T^2,\bar{J}^5)\quad,
\label{eq:rg_cond_T_6}
\end{equation}
where $\bar{Z}=Z(0)$, $\bar{\Gamma}=\Gamma(0)$, $\bar{J}=Z(0)J(0)$, $\bar{J}_I=Z(0)J_I(0)$ and
$\bar{K}=K(0)$. This expression leads us to the final result for the
Fermi liquid coefficient $c_T$ 
\begin{equation}
\label{eq:cT}
c_T\,=\,{\pi^4\over 2}{\bar{J}_I\bar{K}\over\bar{Z}^2}(1+3\bar{J})+O(\bar{J}^5)\,=\,
{\pi^4\bar{J}^3(1+3\bar{J})\over (1-\bar{J})^3}+O(\bar{J}^5),
\end{equation}
where we used $\bar{J}_I=\bar{J}(1-\bar{J})$, $\bar{K}=2\bar{J}^2$ and $\bar{Z}=(1-\bar{J})^2$ [cf.
(\ref{eq:rg_tilde_invariants_TV=0})] in the last step. Comparing $c_T$ with the Fermi liquid
coefficient $c_V$ [cf. (\ref{eq:cV})] we get the familiar universal ratio 
\cite{oguri_JPSJ05,sela_malecki_PRB09} $\frac{c_V}{c_T} = \frac{3}{2 \pi^2}$ of the Fermi liquid 
coefficients in the Kondo model.